\documentclass[aps,prd,amssymb,final,nobibnotes,nofootinbib,twocolumn,superscriptaddress]{revtex4-2}
\usepackage{graphicx}
\usepackage{hyperref}
\usepackage{amsmath}
\usepackage{color}
\usepackage{xcolor}
\usepackage{numprint}
\usepackage{float}
\usepackage{ulem}
\usepackage{physics}
\usepackage{mathtools}
\usepackage{dsfont}
\usepackage{subfigure}
\usepackage{upgreek}
\usepackage{multirow}
\usepackage{mathrsfs}
\usepackage{comment}

\begin{document}
\title{Gravitational bremsstrahlung and the Fulling-Davies-Unruh thermal bath}

\author{Jo\~ao P. B. Brito}
\email{joao.brito@icen.ufpa.br} 
\affiliation{Programa de P\'os-Gradua\c{c}\~{a}o em F\'{\i}sica, Universidade Federal do Par\'a, 66075-110, Bel\'em, Par\'a, Brazil}

\author{Lu\'is C. B. Crispino}
\email{crispino@ufpa.br}
\affiliation{Programa de P\'os-Gradua\c{c}\~{a}o em F\'{\i}sica, Universidade Federal do Par\'a, 66075-110, Bel\'em, Par\'a, Brazil}
		
\author{Atsushi Higuchi}
\email{atsushi.higuchi@york.ac.uk}
\affiliation{Department of Mathematics,  University of York,  Heslington, York YO10 5DD, United Kingdom}

\date{\today}
\begin{abstract}
The electromagnetic radiation emitted by an accelerated charged particle can be described theoretically as the interaction of the charge with the so-called Fulling-Davies-Unruh thermal bath in the coordinate frame co-accelerated with the charge. We present a similar analysis on the gravitational radiation from a classical point mass uniformly accelerated, being pulled by a string satisfying the
weak energy condition.
In particular, we derive the interaction rate (with fixed transverse momentum) of this system of
the point mass and string in the Fulling-Davies-Unruh thermal bath in the co-accelerated frame and show that it equals the graviton emission rate calculated in the standard method in Minkowski spacetime.

\smallskip

\end{abstract}

\maketitle

\section{Introduction}
Following a period of controversies, the classical problem of radiation by a uniformly accelerated charge has become well understood in recent decades. This phenomenon is commonly observed from the perspective of a distant inertial observer, who detects radiation emitted by a charge undergoing an accelerated trajectory relative to that observer (see, e.g., Refs.~\cite{fulton_1960, rohrlich_1961, rohrlich_1963, jackson_1998}). The power observed in an inertial frame can be calculated using the well-known Larmor formula~\cite{larmor_1897}, which has also been generalized to a covariant form~\cite{jackson_1998}. However, the connection of the inertial-frame viewpoint with the viewpoint of a co-accelerated frame of reference, where the charge is static and hence the observer there sees no radiation, remained as an unsolved part of the puzzle. This piece of the puzzle is closely related to the apparent paradox of the equivalence principle~\cite{rohrlich_1963}. In a significant contribution, Boulware investigated this problem and provided an explanation for the apparent paradox of the equivalence principle in Ref.~\cite{boulware_1980}. According to Boulware's findings, the co-accelerated observer sees no radiation because the radiation field can only be unambiguously interpreted in regions of spacetime beyond the observer's horizon. (See also Ref.~\cite{almeida_2006} and the references therein.) 

The inertial and co-accelerated viewpoints about the problem of radiation emitted by a uniformly accelerated charge can also be addressed using a semiclassical framework, i.e., using quantum field theory at tree level, which provides a deeper understanding concerning this problem. In this context, the theoretical description is different in the inertial and accelerated frames, although the response rates of the charge calculated in these frames are the same. While in the inertial frame the source is seen to emit Minkowski quanta, in the co-accelerated frame, or in the Rindler frame, the source is seen to interact with a thermal bath, which is a consequence of the interpretation of the inertial-frame quantum-field state in the accelerated frame. This was shown in Refs.~\cite{higuchi_1992R, higuchi_1992}, where it was concluded that the radiation in the inertial frame is only consistently interpreted in the co-accelerated frame if the Unruh effect~\cite{unruh_1976} is taken into account.
We also note that the Larmor formula was reproduced recently through the modes of photons defined in the Rindler frame using the Unruh effect~\cite{vacalis_2023}.

The Unruh effect is an example of the observer-dependent nature of the field-theoretic particle content~\cite{crispino_2008}. In particular, the Minkowski vacuum, the state with no particles for inertial observers, is described as a thermal bath in the co-accelerated frame of the charge, the Fulling-Davies-Unruh (FDU) thermal bath~\cite{fulling_1973,davies_1975,unruh_1976}. The emission of a Minkowski photon, i.e., a field excitation in the inertial frame, corresponds to either the absorption or emission of Rindler photons, which are the field quanta as viewed by Rindler observers, i.e., by the co-accelerated observers~\cite{unruh_1984}. Thus, in quantum field theory the theoretical description of the acceleration radiation by the inertial observer is associated with the emission of a Minkowski photon by the charge, and the theoretical description by the co-accelerated observer is associated with the emission and absorption of zero-energy Rindler particles, which are shown to be undetectable by co-accelerated observers, agreeing with the classical interpretation~\cite{higuchi_1992,boulware_1980}.

The Unruh effect is an intrinsic effect of quantum field theory and is a necessary ingredient in the description of physics in the Rindler frame. Despite this fact, works questioning the existence of this effect can be found in the literature (see, e.g., Refs.~\cite{belinskii_1997,cruz2016}).
The experimental observation of the FDU thermal bath could dispel any lingering doubt about the Unruh effect. Some proposals have been put forward 
to observe the Unruh effect using state-of-the-art technology, in particular exploiting the above-mentioned description of the radiation using the Unruh effect in the accelerated frame~\cite{cozella_2017} (see also Refs.~\cite{landulfo_2019,akhmedov_2007_1,akhmedov_2007_2}). 
In this proposal, the radiation emitted by a circularly moving charge in the Rindler frame (also interacting with the FDU thermal bath) is compared with the radiation observed in the inertial frame by an experimentalist. The output of the experiment in the inertial frame can be predicted by classical electrodynamics and is found to be the same as inferred by the Rindler observers incorporating the 
Unruh effect. Another example of the quantum phenomenon consistently described by considering the Unruh effect is the decay of uniformly accelerated particles ~\cite{vanzella_2001} (see also Ref.~\cite{suzuki_2003}).

The role played by the Unruh effect in quantum field theory is still a matter of investigation, and the description of many interesting phenomena is lacking. In particular, it will be interesting to confirm that the equality between the response rates of a scalar or an electric charge due to its interaction with the scalar or electromagnetic field, respectively, in the inertial and co-accelerated frames, holds for a point mass interacting with the gravitational field. The main purpose of this paper is to demonstrate in detail that this is indeed the case.

The rest of the paper is organized as follows. In Sec.~\ref{sec:stress-energy} we find a conserved stress-energy tensor, which is described in both inertial and accelerated frames, corresponding to 
a point mass accelerated by a string satisfying the weak energy condition (WEC).  
In Sec.~\ref{sec:perturbations} we review the quantization of gravitational perturbations in Rindler spacetime. In Sec.~\ref{sec:response-rate-Rindler} we calculate the response rate associated with the conserved stress-energy tensor found in Sec.~\ref{sec:stress-energy} in Rindler spacetime, taking into account the Unruh effect.
In Sec.~\ref{sec:response-rate-Minkowski} we calculate the corresponding response rate in the inertial frame. Our final remarks are given in Sec.~\ref{sec:remarks}.  In Appendix~\ref{sec:appendix_C} we show the equality of the response rates computed in the inertial and co-accelerated frames, and in Appendix~\ref{sec:appendix_E} we present the response rate for another stress-energy tensor with a string which does not satisfy the WEC.
We adopt metric signature $(+, -, -, -)$  and units where $8\pi G=c=\hbar=k_B=1.$

\section{Conserved Stress-Energy Tensor}
\label{sec:stress-energy}
Minkowski spacetime is globally covered by the set of coordinates $(t,z,x,y),$ and the right Rindler wedge, which is natural to a uniformly accelerated observer, 
can be covered by Rindler coordinates $(\eta, \xi, x,y),$ 
with these two sets of coordinates related as follows:
\begin{eqnarray}
\label{eq:coord_transform_t}
t &=& a^{-1}e^{a \xi} \sinh a \eta, \\
\label{eq:coord_transform_z}
z &=& a^{-1}e^{a \xi} \cosh a \eta,
\end{eqnarray}
where $z > \abs{t}$ and $-\infty < \eta,\, \xi < \infty.$ The line element is given in 
Rindler coordinates as
\begin{equation}
ds^2 = e^{2 a \xi} \left(d\eta^2-d\xi^2\right)-dx^2-dy^2,
\label{eq:line_element}
\end{equation}
with metric determinant $\mathrm{det}(g_{\mu\nu}) \equiv g  = -e^{4a \xi}.$ The lines of constant $\xi$,
$x,$ and $y$
are the uniformly accelerated trajectories with proper acceleration $a e^{- a \xi}$ and proper time $\tau = e^{a \xi}\eta$ (see, e.g., Ref.~\cite{birrell_1982}). There is a Killing horizon at $z=\abs{t}$ associated with the Killing vector field $\partial_{\eta}.$

Let us first construct a conserved stress-energy tensor in the spacetime described by Eq.~\eqref{eq:line_element} for a system with a uniformly accelerated point mass.  There will be an
external force acting on it because without it the point mass 
would follow a timelike geodesic~\cite{geroch-weatherall-2017}.
We assume that only the components $T^{\eta \eta}$ and $T^{\xi\xi}$ are nonzero and that they are
proportional to $\delta^{(2)}(\mathbf{x}_\bot) = \delta(x)\delta(y)$ with $\mathbf{x}_\bot = (x,y)$ and
otherwise depend only on $\xi$.
Then, the conservation equation, $\nabla_\mu T^{\mu \nu} = 0$, is satisfied if
\begin{equation}
e^{-3 a \xi} \partial_{\xi}\left( e^{ 3 a \xi} T^{\xi \xi}\right) = - a T^{\eta \eta}.
\label{eq:stress_energy_condition}
\end{equation}
We let $T^{\xi \xi}$ be given by
\begin{equation}
T^{\xi \xi} = - \upmu a e^{-2 a \xi} F(\xi) \delta^{(2)}(\mathbf{x}_{\bot}),
\label{eq:solution_Txixi}
\end{equation}
where $\upmu$ is a positive constant and $F(\xi)$ is an arbitrary function. Substituting this equation into Eq.~\eqref{eq:stress_energy_condition}, we find
\begin{equation}
T^{\eta \eta} = \upmu e^{-2 a \xi} \left[F'(\xi) + a F(\xi) \right] \delta^{(2)}(\mathbf{x}_{\bot}).
\label{eq:stress_energy_eta_eta}
\end{equation}
Equations~\eqref{eq:solution_Txixi} and \eqref{eq:stress_energy_eta_eta} satisfy the conservation equation for $T^{\mu\nu}$ for
any $F(\xi)$.  Here, we choose
\begin{equation}
F(\xi) = \theta (\xi),
\label{eq:F}
\end{equation}
where $\theta(x)$ is the Heaviside step function. This choice leads to the following stress-energy tensor:
\begin{eqnarray}
\label{eq:Rindler_T_eta_eta}
T^{\eta \eta} &=& \upmu \left[\delta(\xi) + a e^{- 2 a \xi} \theta (\xi) \right]\delta^{(2)}(\mathbf{x}_{\bot}), \\
\label{eq:Rindler_T_xi_xi}
T^{\xi \xi} &=& -\upmu a e^{-2 a \xi} \theta(\xi)\delta^{(2)}(\mathbf{x}_{\bot}),
\end{eqnarray}
with all the other tensor components vanishing.

\begin{figure}
\center
\includegraphics[scale=0.42]{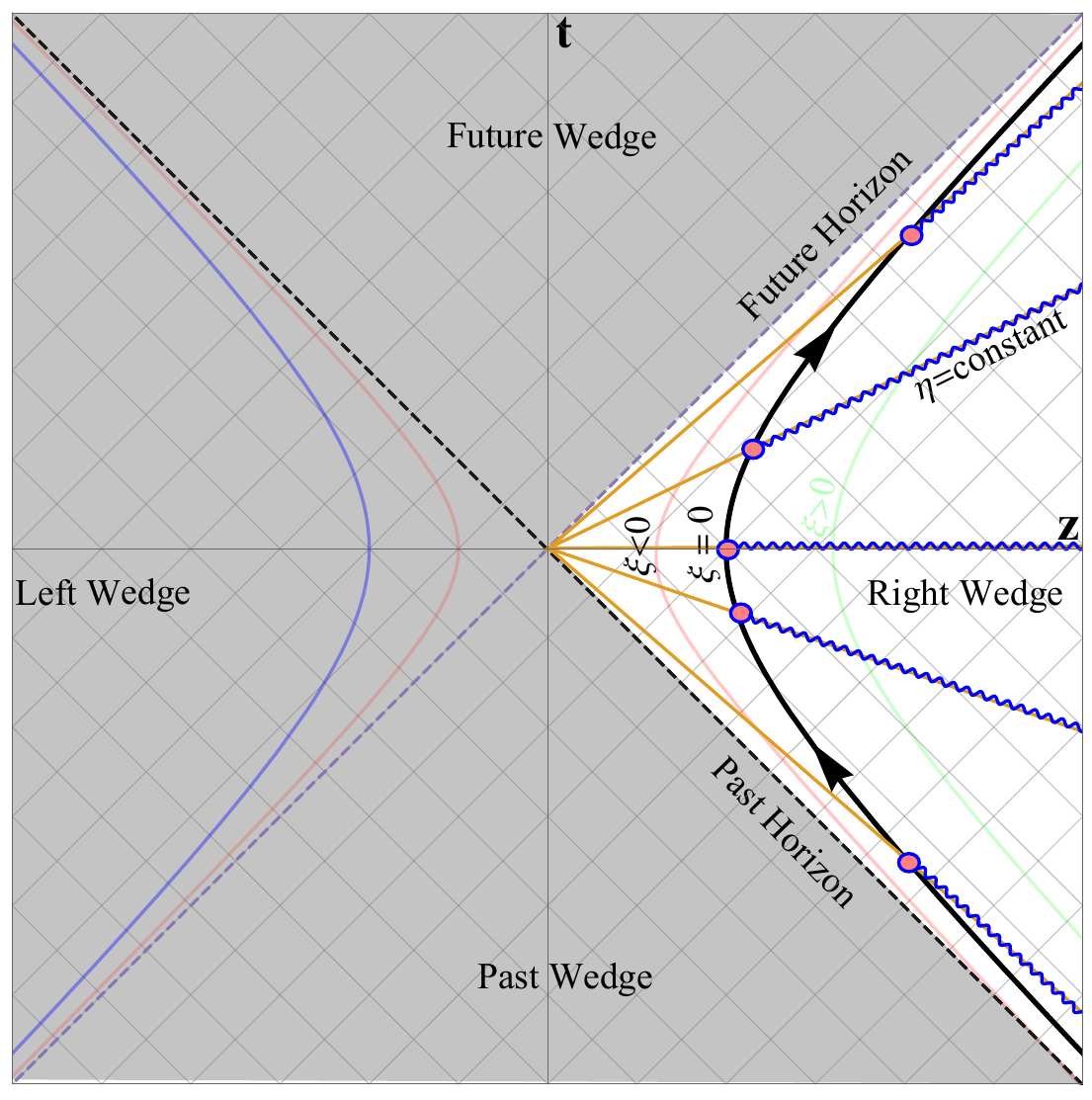}
\caption{Minkowski spacetime diagram and the four Rindler wedges. The trajectory associated with $\xi=0$ is highlighted in the right Rindler wedge. The dots with wavy tails on each $\eta = \mathrm{constant}$ line indicate the location of the source of
the stress-energy tensor at constant Rindler times.}
\label{fig:Rindler_spacetime}
\end{figure}

The stress-energy tensor given by Eqs.~\eqref{eq:Rindler_T_eta_eta} and~\eqref{eq:Rindler_T_xi_xi} describes a point mass $\upmu$ at $(\eta, 0, 0, 0)$ attached to an infinitely long 
string extended along the $\eta = \mathrm{constant}$ line (on the plane $\mathbf{x}_\bot=0$) from
$\xi = 0$ to $\infty$.
This is illustrated in Fig.~\ref{fig:Rindler_spacetime}, where a diagram of Minkowski spacetime is presented.
This stress-energy tensor is given in Minkowski coordinates ($t$ and $z$) 
as
\begin{eqnarray}
\label{eq:Minkowski_T_t_t}
T^{t t} &=& \upmu \left[\cosh^2 a\eta\, \delta( \xi) + a \theta ( \xi) \right]\delta^{(2)}(\mathbf{x}_{\bot}), \\
\label{eq:Minkowski_T_t_z}
T^{t z} &=& \upmu \sinh a\eta \cosh a \eta\, \delta( \xi) \delta^{(2)}(\mathbf{x}_{\bot}), \\
\label{eq:Minkowski_T_z_z}
T^{z z} &=& \upmu \left[\sinh^2 a\eta\, \delta( \xi) - a \theta (\xi) \right]\delta^{(2)}(\mathbf{x}_{\bot}),
\end{eqnarray}
with all other components vanishing.
This stress-energy tensor obeys the WEC.

\section{Gravitational Perturbations and their Quantization in Rindler spacetime}
\label{sec:perturbations}
The Einstein-Hilbert action for pure gravity is
\begin{equation}
    S_{\mathrm{EH}} = -\frac{1}{16\pi G}\int R\,\sqrt{-g^{(f)}}\,d^4 x, \label{eq:Einstein-Hilbert}
\end{equation}
where $R$ is the Ricci scalar curvature, $G$ is Newton's constant, and $g^{(f)}$ is the determinant of the (full) metric tensor $g^{(f)}_{\mu\nu}$. By writing the metric tensor as
\begin{equation}\label{eq:g-h-relation}
    g^{(f)}_{\mu\nu} = g_{\mu\nu}  + \sqrt{32\pi G}\,h_{\mu\nu},
\end{equation}
where $g_{\mu\nu}$ is the flat metric on Minkowski spacetime,
the action~\eqref{eq:Einstein-Hilbert} is given to second order in $h_{\mu\nu}$ as
\begin{equation}
    S_{\textrm{EH}}^{(2)} = \int \mathcal{L}_{\textrm{EH}}^{(2)}\sqrt{-g}\,d^4x,
\end{equation}
where
\begin{eqnarray}
    \mathcal{L}_{\textrm{EH}}^{(2)} & = &  \frac{1}{2}\nabla_\alpha h_{\mu\nu}\nabla^\alpha h^{\mu\nu} - \nabla_\alpha h_{\beta\mu}\nabla^\beta h^{\alpha\mu}\nonumber \\
    & & + \left(\nabla_\alpha h^{\mu\alpha} - \frac{1}{2}\nabla^\mu h\right)\nabla_\mu h, 
\end{eqnarray}
with $h\equiv {h^\alpha}_\alpha$.  Here and below the tensor indices are raised and lowered by the flat metric $g_{\mu\nu}$.  
The Euler-Lagrange equation derived from this Lagrangian is
\begin{eqnarray}
    && \nabla_\alpha\nabla^\alpha h_{\mu\nu} - \nabla_\mu \nabla^\alpha h_{\nu\alpha} - \nabla_\nu \nabla^\alpha h_{\mu\alpha} + \nabla_\mu \nabla_\nu h \nonumber \\
    && \hspace{2cm} +g_{\mu\nu}(\nabla^\alpha\nabla^\beta h_{\alpha\beta} - \nabla_\alpha\nabla^\alpha h) = 0. \label{eq:original-EL}
\end{eqnarray}
This equation is gauge invariant; i.e., a tensor of the form $h_{\mu\nu}^{(G)} = \nabla_\mu \Lambda_\nu + \nabla_\nu\Lambda_\mu$, where $\Lambda_\mu$ is any vector
field, is a solution to this equation.  This gauge invariance allows us to impose the de~Donder condition,
\begin{equation}
    \nabla_\nu h^{\mu\nu} - \frac{1}{2}\nabla^\mu h = 0. \label{eq:de-Donder}
\end{equation}
This condition simplifies the field equation~\eqref{eq:original-EL} to
\begin{equation}
    \nabla_\alpha\nabla^\alpha \left(h_{\mu\nu} - \frac{1}{2}g_{\mu\nu}h\right) = 0. \label{eq:simplified-EL}
\end{equation}

The gravitational perturbations in the right Rindler wedge (and other regions of Minkowski spacetime) were studied in Ref.~\cite{sugiyama_2021}.
In that paper the 
mode functions of positive frequency with respect to the Killing vector $\partial_\eta$ satisfying the de~Donder condition~\eqref{eq:de-Donder} were found in the form
$H_{\mu\nu}(\xi)e^{-i\omega\eta+i\mathbf{k}_\bot\cdot \mathbf{x}_\bot}$, where $\omega$ is a positive constant and $\mathbf{k}_\perp = (k_x,k_y)$ is a constant
two-dimensional vector.  There are two independent nongauge 
modes for given values of $\omega$ and $\mathbf{k}_\bot$, one with even parity and the other with odd parity. 
The odd-parity modes do not couple to the stress-energy tensor $T^{\mu\nu}$ given by Eqs.~\eqref{eq:Rindler_T_eta_eta} and \eqref{eq:Rindler_T_xi_xi} because their
only nonzero components are of the form $h_{\mu\nu}$ with $\mu=\eta,\,\xi$ and $\nu=x,\,y$ (or vice versa). For this reason we discuss only the even-parity modes here.

The even-parity modes can be described in terms of the solutions $\varphi^{(\omega,\mathbf{k}_\bot)}$ to the massless scalar field equation,
\begin{equation}
    \nabla_\alpha\nabla^\alpha \varphi^{(\omega,\mathbf{k}_\bot)} = 0,
\end{equation}
of the form
\begin{equation}
    \varphi^{(\omega,\mathbf{k}_\bot)}(\eta,\xi,\mathbf{x}_\bot) = \phi^{(\omega,k_\bot)}(\xi)e^{-i\omega\eta +i\mathbf{k}_\bot\cdot\mathbf{x}_\bot},
\end{equation}
where $k_\bot = \|\mathbf{k}_\bot\|$.
The functions $\phi^{(\omega,k_\bot)}$ are given by
\begin{equation}
\phi^{(\omega, k_\bot)}(\xi) = \sqrt{\frac{\sinh (\pi \omega/a)}{4 \pi^4 a}} K_{i \omega/a}\left(\frac{k_{\bot} e^{a \xi}}{a} \right),
\label{eq:solution_phi}
\end{equation}
where $K_{\nu}(z)$ is the modified Bessel function of the second kind. The functions $\phi^{(\omega, k_{\bot})}$ are real because
$K_{-\nu}(z)=K_\nu(z)$ and obey the following differential equation:
\begin{equation}
\left[\partial_{\xi}^2 + \left(\omega^2 - k_{\bot}^2 e^{2 a \xi}\right) \right]\phi^{(\omega, k_{\bot})}(\xi)=0.
\label{eq:dif_phi}
\end{equation}
The mode functions $\varphi^{(\omega,\mathbf{k}_\bot)}$ satisfy the Klein-Gordon normalization condition~\cite{fulling_1989},
\begin{eqnarray}
 \langle\varphi^{(\omega,\mathbf{k}_\bot)},\varphi^{(\omega',\mathbf{k}'_\bot)}\rangle_{\textrm{sc}}  & \equiv &  i\int_{\Sigma} \overline{\varphi^{(\omega,\mathbf{k}_\bot)}}\stackrel{\longleftrightarrow}{\nabla_\mu} 
    \varphi^{(\omega',\mathbf{k}_\bot')}n^\mu\,d\Sigma \nonumber \\
    & = & \delta(\omega-\omega')\delta^{(2)}(\mathbf{k}_\bot - \mathbf{k}'_\bot), \label{eq:KG-inner-product}
\end{eqnarray}
where $\Sigma$ is the $\eta$\,=\,constant hypersurface for the right Rindler wedge and where $n^\mu$ is the future-pointing unit normal to this hypersurface.

The even-parity modes are given in the form
\begin{equation}\label{eq:even-modes}
 h_{\mu\nu}^{(\omega,\mathbf{k}_\bot)}(\eta,\xi,\mathbf{x}_\bot)
 = H_{\mu\nu}^{(\omega,k_\bot)}(\xi)e^{-i\omega\eta+i\mathbf{k}_\bot\cdot \mathbf{x}_\bot},
\end{equation}
where 
the nonzero components of $H_{\mu\nu}^{(\omega,k_\bot)}(\xi)$ are~\cite{sugiyama_2021}
\begin{eqnarray}
H^{(\omega,k_\bot)}_{\eta \eta}(\xi) &  = & H^{(\omega,k_\bot)}_{\xi \xi}(\xi) \nonumber \\
& = & \frac{1}{\sqrt{2}}\left[-e^{2 a \xi} + \frac{2}{k_{\perp}^2} \left( \omega^2 + a \partial_{\xi} \right)\right]\phi^{(\omega,k_{\bot})}(\xi),
\nonumber\\ \label{eq:SYK_normalized_modes_eta_eta_xi_xi}\\
\label{eq:SYK_normalized_modes_eta_xi}
H_{\eta \xi}^{(\omega,k_\bot)}(\xi) &=& \frac{\sqrt{2} i \omega}{k_{\perp}^2}\left(\partial_{\xi}-a \right) \phi^{(\omega, k_{\bot})}(\xi), \\
\label{eq:SYK_normalized_modes_i_j}
H_{i j}^{(\omega,k_\bot)}(\xi) &=& -\frac{\delta_{i j}}{\sqrt{2}}\phi^{(\omega,k_{\bot})}(\xi),
\end{eqnarray}
where $i$ and $j$ are $x$ or $y$.

To discuss the normalization condition for these modes it will be useful to consider physically equivalent mode functions related to them by a gauge transformation.
We define the vector $\Lambda_\mu^{(\omega,\mathbf{k}_\bot)}$ by
\begin{eqnarray}
    \Lambda^{(\omega,\mathbf{k}_\bot)}_\eta(\eta,\xi,\mathbf{x}_\bot) & = & \frac{1}{\sqrt{2}\,k_\perp^2}\partial_\eta \varphi^{(\omega,\mathbf{k}_\bot)}(\eta,\xi,\mathbf{x}_\bot),\\
    \Lambda^{(\omega,\mathbf{k}_\bot)}_\xi(\eta,\xi,\mathbf{x}_\bot) & =  & \frac{1}{\sqrt{2}\,k_\perp^2}\partial_\xi\varphi^{(\omega,\mathbf{k}_\bot)}(\eta,\xi,\mathbf{x}_\bot),\\
    \Lambda^{(\omega,\mathbf{k}_\bot)}_i(\eta,\xi,\mathbf{x}_\bot) & =  & -\frac{1}{\sqrt{2}\,k_\perp^2}\partial_i\varphi^{(\omega,\mathbf{k}_\bot)}(\eta,\xi,\mathbf{x}_\bot),
\end{eqnarray}
where $i=x,y$.  We also define the gauge-transformed mode functions by
\begin{eqnarray}
    \widetilde{h}_{\mu\nu}^{(\omega,\mathbf{k}_\bot)} & = & h_{\mu\nu}^{(\omega,\mathbf{k}_\bot)} + \nabla_\mu \Lambda^{(\omega,\mathbf{k}_\bot)}_\nu
    + \nabla_\nu \Lambda^{(\omega,\mathbf{k}_\bot)}_\mu.\label{eq:gauge-transformation}
\end{eqnarray}
Noting that the nonzero Christoffel symbols are $\Gamma^\eta_{\eta\xi} = \Gamma^\eta_{\xi\eta} = \Gamma^\xi_{\eta\eta} = \Gamma^\xi_{\xi\xi} = a$, we readily find
\begin{equation}
 \widetilde{h}^{(\omega,\mathbf{k}_\bot)}_{\mu\nu} = -\frac{1}{\sqrt{2}}\left[ g_{\mu\nu} + 2q_{\mu\nu}(\mathbf{k}_\bot)\right]
 \varphi^{(\omega,\mathbf{k}_\bot)}, \label{new-field}
\end{equation}
where
\begin{equation}
q_{\mu\nu}(\mathbf{k}_\bot)  = \begin{cases}
\delta_{ij} - \frac{k_ik_j}{k_\perp^2} & \textrm{if}\ \mu,\nu=x\ \textrm{or}\ y,\\
 0 & \textrm{otherwise}.
\end{cases} \label{new-field-transverse}
\end{equation}
It is straightforward to show that the mode functions $\widetilde{h}_{\mu\nu}^{(\omega,\mathbf{k}_\bot)}$ satisfy both the de~Donder condition~\eqref{eq:de-Donder}
and the field equation~\eqref{eq:simplified-EL}.

The mode functions describing the gravitational perturbations are normalized with respect to the inner product analogous to the 
Klein-Gordon inner product~\eqref{eq:KG-inner-product}.  We first define the conjugate momentum current for a solution $h^{(i)}_{\mu\nu}$ of the
Euler-Lagrange equation~\eqref{eq:original-EL} as follows (see, e.g., Ref.~\cite{faizal_2012}):
\begin{eqnarray}
    p_{(i)}^{\alpha\mu\nu} & = & \left.\frac{\partial\mathcal{L}^{(2)}_{\textrm{EH}}}{\partial \nabla_\alpha h_{\mu\nu}}\right|_{h=h^{(i)}}.
\end{eqnarray}
Then the Euler-Lagrange equation~\eqref{eq:original-EL} can be written as
\begin{equation}
    \nabla_\alpha p_{(i)}^{\alpha\mu\nu}  - \left.\frac{\partial \mathcal{L}^{(2)}_{\textrm{EH}}}{\partial h_{\mu\nu}}\right|_{h=h^{(i)}} = 0. \label{eq:EL-canonical}
\end{equation}
(In our case, i.e., in flat spacetime, the second term is absent.)  Since the Lagrangian density $\mathcal{L}_{\textrm{EH}}^{(2)}$ is quadratic in $h_{\mu\nu}$,
we have, for any two solutions $h^{(i)}_{\mu\nu}$ and $h^{(j)}_{\mu\nu}$ to Eq.~\eqref{eq:original-EL}, 
\begin{equation}
\nabla_\alpha \overline{h^{(i)}_{\mu\nu}}p_{(j)}^{\alpha\mu\nu} = \overline{p_{(i)}^{\alpha\mu\nu}}\nabla_\alpha h^{(j)}_{\mu\nu}. 
\end{equation}
This equation and Eq.~\eqref{eq:EL-canonical} imply that the following inner product is $\eta$-independent:
\begin{equation}
    \langle h^{(i)},h^{(j)}\rangle_{\textrm{ts}} \equiv i \int_{\Sigma} \left( \overline{h^{(i)}_{\mu\nu}}p^{\alpha\mu\nu}_{(j)}
    - \overline{p_{(i)}^{\alpha\mu\nu}}h^{(j)}_{\mu\nu}\right)n_\alpha d\Sigma,
\end{equation}
where $\Sigma$ and $n^\mu$ are defined as in Eq.~\eqref{eq:KG-inner-product}.  If the solutions $h_{\mu\nu}^{(i)}$ and $h_{\mu\nu}^{(j)}$ obey the de~Donder
condition~\eqref{eq:de-Donder}, then this inner product simplifies to
\begin{eqnarray}
    && \langle h^{(i)},h^{(j)}\rangle_{\textrm{dD}}\nonumber \\
    && = i\int_{\Sigma}\left( \overline{h^{(i)}_{\mu\nu}}\stackrel{\longleftrightarrow}{\nabla_\alpha} h^{(j)\mu\nu}
    - \frac{1}{2}\overline{h^{(i)}}\stackrel{\longleftrightarrow}{\nabla_\alpha} h^{(j)}\right)n^\alpha\,d\Sigma. \nonumber \\
\end{eqnarray}
For the mode functions~\eqref{new-field} we find
\begin{eqnarray}
    \langle \widetilde{h}^{(\omega,\mathbf{k}_\bot)},\widetilde{h}^{(\omega',\mathbf{k}_\bot')}\rangle_{\textrm{dD}}
    & = & \langle \varphi^{(\omega,\mathbf{k}_\bot)},\varphi^{(\omega',\mathbf{k}_\bot')}\rangle_{\textrm{sc}} \notag \\
    & = & \delta(\omega-\omega')\delta^{(2)}(\mathbf{k}_\bot - \mathbf{k}'_\bot). \label{eq:inner-product-even-modes}
\end{eqnarray}

Now let us discuss the quantization of the even-parity metric perturbations,
which couple to the stress-energy tensor $T^{\mu\nu}$ 
in Eqs.~\eqref{eq:Rindler_T_eta_eta} and \eqref{eq:Rindler_T_xi_xi}.  After fixing the gauge completely, the quantized even-parity gravitation field can be expressed as
\begin{eqnarray}
    \widehat{h}_{\mu\nu}^{(\textrm{even})} (\eta,\xi,\mathbf{x}_\bot)
    & = & \int_{0}^\infty d\omega \int d^2\mathbf{k}_\bot \nonumber \\
    & & \times \left[ \widetilde{h}_{\mu\nu}^{(\omega,\mathbf{k}_\bot)} (\eta,\xi,\mathbf{x}_\bot)\hat{a}_{(\omega,\mathbf{k}_\bot)} + \textrm{H.c.} \right].
    \nonumber \\ \label{eq:tensor_mode_operator}
\end{eqnarray}
As in the scalar case~\cite{wald-QFT-textbook}, the inner product~\eqref{eq:inner-product-even-modes} implies
\begin{equation}
    [ \hat{a}_{(\omega,\mathbf{k}_\bot)}, \hat{a}^\dagger_{(\omega',\mathbf{k}_\bot')}] = \delta(\omega-\omega')\delta^{(2)}(\mathbf{k}_\bot - \mathbf{k}'_\bot).
\end{equation}
The Fulling vacuum state $\ket{0_{\mathrm{F}}}$~\cite{fulling_1973}, which is the no-particle state in the Rindler frame, satisfies $\hat{a}_{(\omega,\mathbf{k}_\bot)}\ket{0_{\mathrm{F}}} = 0$ for all $\omega$ and $\mathbf{k}_\bot$.

\section{Response Rate in the Accelerated Frame}
\label{sec:response-rate-Rindler}

The stress-energy tensor is defined from a matter action $S_{\textrm{matter}}$ by
\begin{equation}
    T^{\mu\nu} = \frac{2}{\sqrt{-g^{(f)}}}\frac{\delta S_{\textrm{matter}}}{\delta g^{(f)}_{\mu\nu}}.
\end{equation}
At linear order we find from Eq.~\eqref{eq:g-h-relation}
\begin{equation}
    \left.\frac{\delta S_{\textrm{matter}}}{\delta h_{\mu\nu}}\right|_{h=0} = \sqrt{8\pi G}\,T^{\mu\nu}.
\end{equation}
Hence, the interaction term describing the one-graviton processes due to the classical stress-energy tensor $T^{\mu\nu}$ given by Eqs.~\eqref{eq:Rindler_T_eta_eta} and 
\eqref{eq:Rindler_T_xi_xi} can be taken to be
\begin{equation}
\widehat{S}_{\mathrm{int}} = \sqrt{8\pi G}\int T^{\mu \nu}\widehat{h}^{(\textrm{even})}_{\mu \nu}\sqrt{-g}\,d^4x,
\label{eq:int_action}
\end{equation}
where the operator $\widehat{h}^{(\textrm{even})}_{\mu \nu}(x)$ is given by Eq.~\eqref{eq:tensor_mode_operator}.
We set $8\pi G=1$ from now on.

The probability amplitudes associated with the emission and absorption of a graviton with frequency $\omega$ and transverse momentum $\mathbf{k}_{\bot}$ in the Fulling vacuum state 
$\ket{0_{\mathrm{F}}}$ are given by
\begin{eqnarray}
\mathcal{A}_{\mathrm{em}}^{(\omega,\mathbf{k}_{\perp})} &=& \bra{0_{\mathrm{F}}}\hat{a}_{(\omega,\mathbf{k}_\bot)}  \widehat{S}_{\mathrm{int}} \ket{0_{\mathrm{F}}} \nonumber\\
\label{eq:amplitude_R_emission2}
&=& \int T^{\mu \nu} \overline{h^{(\omega,\mathbf{k}_\bot)}_{\mu \nu}}\sqrt{-g}\,d^4 x, \\
\mathcal{A}_{\mathrm{abs}}^{(\omega,\mathbf{k}_{\perp})} &=& \bra{0_{\mathrm{F}}} \widehat{S}_{\mathrm{int}} \hat{a}^\dagger_{(\omega,\mathbf{k}_\bot)}\ket{0_{\mathrm{F}}}\nonumber\\
&=& \int T^{\mu \nu} h^{(\omega,\mathbf{k}_\bot)}_{\mu \nu}\sqrt{-g}\,d^4x,
\label{eq:amplitude_R_absorption2}
\end{eqnarray}
respectively, where $h^{(\omega,\mathbf{k}_\bot)}_{\mu\nu}$ is
the even-parity positive-frequency perturbation with frequency $\omega$ and transverse momentum $\mathbf{k}_\perp$
given by Eq.~\eqref{eq:even-modes}.  Here, we replaced the gauge-transformed mode functions $\widetilde{h}^{(\omega,\mathbf{k}_\bot)}_{\mu\nu}$,
which are used in the expansion of the even-parity quantum graviton 
field $\widehat{h}^{(\textrm{even})}_{\mu\nu}$ in Eq.~\eqref{eq:tensor_mode_operator}, with the original mode functions
$h^{(\omega,\mathbf{k}_\bot)}_{\mu\nu}$ without changing the result because the stress-energy tensor satisfies
$\nabla_\mu T^{\mu\nu} = 0$.
The differential \textit{spontaneous} emission probability, i.e., the probability of emission of one graviton with fixed transverse momentum $\mathbf{k}_{\perp}$ and Rindler energy $\omega$, is given by
\begin{equation}
d\mathcal{P}_{\omega,\mathbf{k}_{\bot}}^{\mathrm{em}} = \abs{\mathcal{A}_{\mathrm{em}}^{(\omega,\mathbf{k}_\perp)}}^2 d\omega d^2\mathbf{k}_\bot.
\label{eq:probability_R_emission}
\end{equation}

The Minkowski vacuum state is equivalent to a thermal bath of the Unruh temperature $T_U = a/2\pi$ for processes confined to the
right Rindler wedge~\cite{unruh_1976}.  
This correspondence exemplifies the observer-dependent notion of the field-theoretic particle content. Here, the notion of a particle is defined 
with respect to the timelike Killing vector $\partial_t$ for the inertial observers and 
with respect to $\partial_{\eta}$ for the co-accelerated observers~\cite{fulling_1973}. 
The expected number of gravitons
with a given (normalized) wave function with (approximate) frequency $\omega$ that surround the co-accelerated observers is
\begin{equation}
n(\omega) \equiv \frac{1}{e^{2\pi\omega/a}-1},
\label{eq:bath}
\end{equation}
which is the Bose-Einstein distribution function with the Unruh temperature $a/2\pi$.
Thus, in the Rindler frame there is induced emission as well as spontaneous emission.  If we write the total emission rate as
\begin{align}
    \mathcal{R}^{\mathrm{em}}_{\mathrm{tot}} = \int  \mathcal{R}^{\textrm{em}}_{\mathbf{k}_\perp}\,d^2\mathbf{k}_\bot,
\end{align}
then the differential emission rate, $\mathcal{R}^{\mathrm{em}}_{\mathbf{k}_\perp}$, i.e., the emission rate per unit time and transverse momentum squared, is given by
\begin{align}
\label{eq:emission_rate_Rindler}
    \mathcal{R}^{\mathrm{em}}_{\mathbf{k}_\perp} = \frac{1}{T_0}\int_0^\infty 
    \abs{\mathcal{A}_{\mathrm{em}}^{(\omega,\mathbf{k}_\perp)}}^2[1+n(\omega)]d\omega ,
\end{align}
where $T_0 = \int_{-\infty}^\infty d\eta$ is the ``infinite total proper time.''\footnote{We present a derivation of the result in this section without the use of
infinite proper time in Appendix~\ref{sec:appendix_C}.} Equation~\eqref{eq:emission_rate_Rindler} was obtained from Eq.~\eqref{eq:probability_R_emission} taking into account the spontaneous and induced emission due to the thermal bath associated with $n(\omega)$. The differential absorption rate, i.e., the absorption rate per unit time and transverse momentum squared, is
\begin{align}
    \mathcal{R}^{\mathrm{abs}}_{\mathbf{k}_\perp} = \frac{1}{T_0}\int_0^\infty
    \abs{\mathcal{A}_{\mathrm{em}}^{(\omega,\mathbf{k}_\perp)}}^2n(\omega) d\omega,
\end{align}
where we used the relation $\mathcal{A}_{\mathrm{abs}}^{(\omega,\mathbf{k}_\bot)} = \overline{\mathcal{A}_{\mathrm{em}}^{(\omega,-\mathbf{k}_\bot)}}$,
which follows from Eqs.~\eqref{eq:amplitude_R_emission2} and \eqref{eq:amplitude_R_absorption2}, and Eq.~\eqref{eq:probability_R_emission} taking into account the thermal factor $n(\omega)$.
The differential \textit{interaction} rate, $\mathcal{R}_{\mathbf{k}_\perp}$, 
is the sum of the differential emission and absorption rates, 
$\mathcal{R}^{\mathrm{em}}_{\mathbf{k}_\perp}$ and 
$\mathcal{R}^{\mathrm{abs}}_{-\mathbf{k}_\perp}$.  Noting that
 $n(-\omega) = -[1+n(\omega)]$, we find
\begin{align}\label{eq:diff-interaction-rate}
    \mathcal{R}_{\mathbf{k}_\perp} = \frac{1}{T_0}\int_{-\infty}^\infty  
    \abs{\mathcal{A}_{\mathrm{em}}^{(\omega,\mathbf{k}_\perp)}}^2|1+n(\omega)|d\omega.
\end{align}

As we will see, Eq.~\eqref{eq:diff-interaction-rate}, which follows directly from the perturbation theory in Rindler spacetime taking into account the FDU thermal bath, gives the same result as the emission rate obtained using standard quantum field theory in Minkowski spacetime.

To find the amplitude $\mathcal{A}_{\mathrm{em}}^{(\omega,\mathbf{k}_\perp)}$ we first find from Eqs.~\eqref{eq:Rindler_T_eta_eta} and
\eqref{eq:Rindler_T_xi_xi}, noting that $\sqrt{-g}=e^{2a\xi}$,
\begin{eqnarray}
  && \int T^{\eta \eta} \overline{h_{\eta \eta}^{(\omega,\mathbf{k}_\bot)}}\sqrt{-g}\,d^4x \nonumber \\
 &&= \upmu I(\omega) \left[ \overline{H^{(\omega,k_\bot)}_{\eta \eta}(0)} + a \int_{0}^{\infty} \overline{H^{(\omega,k_\bot)}_{\eta \eta}(\xi)}\,d\xi \right], \label{eq:integrand_eta} \\
&& \int T^{\xi \xi} \overline{h^{(\omega,\mathbf{k}_\bot)}_{\xi \xi}}\sqrt{-g}\,d^4x \nonumber \\
&& = -\upmu a I(\omega) \int_{0}^{\infty} \overline{H_{\xi \xi}^{(\omega,k_\bot)}(\xi)}\,d\xi,
\label{eq:integrand_xi}
\end{eqnarray}
where the functions $H_{\eta\eta}^{(\omega,k_\bot)}$ and $H_{\xi\xi}^{(\omega,k_\bot)}$ are defined by Eqs.~\eqref{eq:even-modes} and \eqref{eq:SYK_normalized_modes_eta_eta_xi_xi} and where
\begin{equation}
    I(\omega) = \int_{-\infty}^\infty e^{i\omega\eta}\,d\eta. \label{eq:infinite-delta}
\end{equation}
The first (second) term of the right-hand side of Eq.~\eqref{eq:integrand_eta} comes from the point mass (extended string). 
Since $H^{(\omega,k_\bot)}_{\eta\eta}(\xi)=H^{(\omega,k_\bot)}_{\xi\xi}(\xi)$, we obtain
\begin{eqnarray}
\mathcal{A}_{\mathrm{em}}^{(\omega,\mathbf{k}_\perp)} &  = & \int T^{\mu \nu} \overline{h^{(\omega,\mathbf{k}_\bot)}_{\mu \nu}}\sqrt{-g}\,d^4x
\label{eq:referred-to-in-C} \\
& = &  \upmu I(\omega) \overline{H^{(\omega,k_\bot)}_{\eta \eta}(0)}.
\label{eq:integrand_of_A}
\end{eqnarray}
Thus, the contribution of the string to the probability amplitude vanishes, and only the point mass contributes. However, 
this vanishing contribution of the string is gauge specific, and the string contribution is nonzero if we use the gauge-transformed modes $\widetilde{h}_{\mu\nu}^{(\omega,\mathbf{k}_\bot)}$ given by Eq.~\eqref{new-field} as we show below.

The differential interaction rate, $\mathcal{R}_{\mathbf{k}_\perp}$, is found by substituting Eqs.~\eqref{eq:bath} and 
\eqref{eq:integrand_of_A} with \eqref{eq:infinite-delta} into Eq.~\eqref{eq:diff-interaction-rate}.  In squaring
the amplitude $\mathcal{A}_{\mathrm{em}}^{(\omega,\mathbf{k}_\perp)}$ given by Eq.~\eqref{eq:integrand_of_A} we employ the
standard formal manipulation $|I(\omega)|^2 = 2\pi T_0 \delta(\omega)$.  Thus, we find
\begin{equation}
\mathcal{R}_{\mathbf{k}_{\bot}} = \left[ 2\pi \upmu^2 \abs{ H^{(\omega,k_\bot)}_{\eta \eta}(0)}^2[1+n(\omega)]\right]_{\omega\to 0}.
\label{eq:induced_emission_expli}
\end{equation}
Note that in the limit $\omega \to 0$ we have $n(\omega) \to \infty$: the Bose-Einstein distribution function diverges for the low-Rindler-energy gravitons as
\begin{equation}
    n(\omega) = \frac{a}{2\pi\omega} + O(1), \label{eq:bath-small-omega}
\end{equation}
for small $\omega$.
On the other hand, Eq.~\eqref{eq:SYK_normalized_modes_eta_eta_xi_xi} with Eq.~\eqref{eq:solution_phi} leads to
\begin{align}
H^{(\omega,k_\bot)}_{\eta\eta}(0)
& = -\sqrt{\frac{\pi \omega}{8 \pi^4 a^2}}\notag \\
& \quad\times\left[ K_{0}\left(\frac{k_\perp}{a}\right) 
- \frac{2a}{k_\bot}K_0'\left(\frac{k_\bot}{a}\right)
+ O(\omega)\right],\label{eq:h-eta-eta-0}
\end{align}
for small $\omega$.
Thus, we have
$H^{(\omega,k_\bot)}_{\eta\eta}(0)\to 0$ as $\omega\to 0$.  This reflects the fact that the stress-energy tensor is static in the Rindler frame,
and, hence, the spontaneous emission rate vanishes.  However, the induced emission rate is nonzero 
because of the interaction of the source with the zero-energy 
Rindler gravitons from the thermal bath, whose state density is infinite. 
We readily find the interaction rate by substituting Eqs.~\eqref{eq:bath-small-omega} and 
\eqref{eq:h-eta-eta-0} into Eq.~\eqref{eq:induced_emission_expli} as
\begin{equation}
\mathcal{R}_{\mathbf{k}_{\bot}} = \frac{\upmu^2}{8 \pi^3 a} \abs{K_{2}\left(\frac{k_{\bot}}{a}\right)}^2,
\label{eq:response_rate}
\end{equation}
where we have used the equality,
\begin{equation}\label{eq:K0-K2-equation}
    K_0(z)- \frac{2}{z}K'_0(z) = K_2(z),
\end{equation}
which can be derived from Ref.~\cite[\S 10.29(i)]{NIST_handbook}.
In Appendix~\ref{sec:appendix_C}
we present a derivation of this result without the use of infinite interaction time.

We should obtain the same result if 
we use the mode functions $\widetilde{h}^{(\omega,\mathbf{k}_\bot)}_{\mu\nu}$ 
given by Eq.~\eqref{new-field}
instead of $h^{(\omega,\mathbf{k}_\bot)}_{\mu\nu}$ because these modes are related by a gauge transformation
[see Eq.~\eqref{eq:gauge-transformation}]. Let us verify this fact.  By expressing the mode functions
$\widetilde{h}_{\mu\nu}^{(\omega,\mathbf{k}_\bot)}$ as
\begin{equation}
    \widetilde{h}_{\mu\nu}^{(\omega,\mathbf{k}_\bot)}(\eta,\xi,\mathbf{x}_\bot) = 
    \widetilde{H}_{\mu\nu}^{(\omega,k_\bot)}(\xi) e^{-i\omega\eta + i\mathbf{k}_\bot\cdot\mathbf{x}_\bot},
\end{equation}
we find
\begin{eqnarray}
    \widetilde{H}_{\eta\eta}^{(\omega,k_\bot)}(\xi)
    & = & - \widetilde{H}_{\xi\xi}^{(\omega,k_\bot)}(\xi) \\
    & = & - \frac{1}{\sqrt{2}}e^{2a\xi}\phi^{(\omega,k_\bot)}(\xi),
\end{eqnarray}
where the functions $\phi^{(\omega,k_\bot)}$ are given by Eq.~\eqref{eq:solution_phi}.
Then, Eq.~\eqref{eq:integrand_of_A} is replaced by 
\begin{align}
& \int T^{\mu \nu} \overline{\widetilde{h}^{(\omega,\mathbf{k}_\bot)}_{\mu \nu}}\sqrt{-g}\,d^4x
\notag \\
& = \upmu I(\omega) \left[ \overline{\widetilde{H}_{\eta \eta}(0)} + 2 a \int_{0}^{\infty} \overline{\widetilde{H}_{\eta \eta}(\xi)}\,d\xi\right] \\
& = - \upmu I(\omega) \sqrt{\frac{\pi \omega}{8 \pi^4 a^2}}\notag \\
& \quad\times\left[ K_{0}\left(\frac{k_\perp}{a}\right) 
+ \frac{2a^2}{k_\bot^2}\int_{k_\bot/a}^\infty y K_0(y)dy
+ O(\omega)\right], \label{eq:integrand_of_A_gauge}
\end{align}
where we have changed the integration variable from $\xi$ to
$y=(k_\perp/a)e^{a\xi}$, for small $\omega$.
Notice that the contribution from the string does not vanish.  By comparing this formula with
Eq.~\eqref{eq:integrand_of_A}, where $H_{\eta\eta}(0)$ is given by 
Eq.~\eqref{eq:h-eta-eta-0}, one finds that the equality of the response rates follows if
\begin{align}
    z K_0'(z) = - \int_z^{\infty} yK_0(y)dy. 
\end{align}
Since both sides tend to $0$ as $z\to\infty$ because $K_{\nu}(z) \approx \sqrt{\pi/2z}\,e^{-z}$
for large $z$, this equality is equivalent to its derivative, i.e.,
\begin{align}\label{eq:diff-eq-for-K0}
    zK_0''(z) + K_0'(z) = zK_0(z),
\end{align}
which is the modified Bessel equation satisfied by $K_0(z)$.  Thus, we have verified that the mode 
functions 
$\widetilde{h}_{\mu\nu}^{(\omega,\mathbf{k}_\bot)}$ also lead to the differential interaction rate
$\mathcal{R}_{\mathbf{k}_\bot}$ given by Eq.~\eqref{eq:response_rate}.

\section{Response Rate in the Inertial Frame}
\label{sec:response-rate-Minkowski}
The spontaneous emission probability in Minkowski spacetime for fixed transverse momentum $\mathbf{k}_{\bot}$ for any conserved stress-energy tensor $T^{\mu\nu}$ given by Eqs.~\eqref{eq:Minkowski_T_t_t}--\eqref{eq:Minkowski_T_z_z}
can be written as
\begin{eqnarray}
\label{eq:transition_probability_Minkowski1}
\mathcal{P}_{\mathbf{k}_{\bot}}^{\mathrm{M}} &=& \int \frac{dk_z}{(2\pi)^3 2 k} \left( \overline{\mathcal{T}^{\mu \nu}}\mathcal{T}_{\mu \nu} - \frac{1}{2}\overline{\mathcal{T}}\mathcal{T}\right) \\
&=&\int \frac{dk_z}{(2\pi)^3 2 k} \left(\frac{1}{2}\abs{\mathcal{T}^{t t}+\mathcal{T}^{z z}}^2 - 2 \abs{\mathcal{T}^{t z}}^2\right), \nonumber \\
\label{eq:transition_probability_Minkowski2}
\end{eqnarray}
where $k \equiv \norm{\mathbf{k}}\equiv \norm{(k_z,\mathbf{k}_{\bot})}$ and $\mathcal{T}^{\mu \nu}(\mathbf{k})$ is defined as
\begin{equation}
\mathcal{T}^{\mu \nu}(\mathbf{k}) = \int e^{i (k t - \mathbf{k} \cdot \mathbf{x})} T^{\mu \nu}(x)\,d^4x,
\label{eq:fourier_T}
\end{equation}
and $\mathcal{T}(\mathbf{k}) \equiv \mathcal{T}^{\mu}{}_{\mu}(\mathbf{k})$.  
The total emission probability would be the integral of $\mathcal{P}_{\mathbf{k}_{\bot}}^{\mathrm{M}}$ 
over $\mathbf{k}_\bot$.

Equation~\eqref{eq:transition_probability_Minkowski1} can be derived from the well-known expression for the total emission probability,
\begin{equation}
\label{eq:em_prob_Mink0}
    \mathcal{P}^{M} = \int \frac{d^3\mathbf{k}}{(2\pi)^3 2k} \,  \overline{\mathcal{T}^{\mu \nu}(k)} \sum_{\lambda} \epsilon_{\mu \nu}^{\lambda} \epsilon_{\alpha \beta}^{\lambda} \mathcal{T}^{\alpha \beta}(k) \,,
\end{equation}
where  $\epsilon_{\mu \nu}^{\lambda}$ is the Minkowski polarization matrix, with $\lambda$ denoting the independent polarization states, by using $k_{\nu}\mathcal{T}^{\mu \nu}=0$ and choosing, for simplicity, a coordinate system such that $k^{\mu}=(k,0,0,k),$ with
\begin{equation}
\epsilon_{1 1}^{\lambda_1}=-\epsilon_{2 2}^{\lambda_1}=\epsilon_{1 2}^{\lambda_2}=\epsilon_{2 1}^{\lambda_2}=\frac{1}{\sqrt{2}},
\end{equation}
and the other components vanishing.

To find the components of $\mathcal{T}^{\mu \nu}(\mathbf{k})$ it is convenient to change the integration variables
from $(t,z)$ to $(\eta, \xi)$ using Eqs.~\eqref{eq:coord_transform_t} and~\eqref{eq:coord_transform_z}.
Thus, we find
\begin{equation}
\mathcal{T}^{\mu \nu}(\mathbf{k}) = \int_{-\infty}^\infty \left[\int_{-\infty}^\infty e^{2 a \xi} \exp \left( i \frac{e^{a\xi}}{a} \mathcal{K} \right)T^{\mu \nu}_{(\mathrm{R})}(\eta,\xi)d\xi\right]d\eta,
\label{eq:fourier_T_coord_eta_xi}
\end{equation}
where $\mathcal{K}\equiv  k \sinh a\eta - k_z \cosh a\eta$ and
\begin{equation}
    T^{\mu\nu}(x)= T^{\mu\nu}_{(\mathrm{R})}(\eta,\xi)\delta^{(2)}(\mathbf{x}_\bot).
\end{equation}
Then, using Eqs.~\eqref{eq:Minkowski_T_t_t}--\eqref{eq:Minkowski_T_z_z}, we obtain
\begin{widetext}
\begin{eqnarray}
\label{eq:fourier_T_t_t}
\mathcal{T}^{t t}(\mathbf{k}) &=& \upmu \int_{-\infty}^\infty  \left\{ \cosh^2 a\eta \exp \left( \frac{i}{a}\mathcal{K} \right) + a \int_{0}^{\infty} d\xi e^{2 a \xi} \exp \left(i \frac{e^{a\xi}}{a} \mathcal{K} \right) \right\}d\eta, \\
\mathcal{T}^{z z}(\mathbf{k}) &=& \upmu \int_{-\infty}^\infty\left\{ \sinh^2 a\eta \exp \left( \frac{i}{a}\mathcal{K} \right) - a \int_{0}^{\infty} d\xi e^{2 a \xi} \exp \left(i \frac{e^{a\xi}}{a} \mathcal{K} \right) \right\}d\eta.
\label{eq:fourier_T_z_z}
\end{eqnarray}
Note that the first (second) term of the right-hand side of Eqs.~\eqref{eq:fourier_T_t_t} and~\eqref{eq:fourier_T_z_z} comes from the point mass
(extended string). By adding Eqs.~\eqref{eq:fourier_T_t_t} and~\eqref{eq:fourier_T_z_z}, we find  
\begin{equation}
\mathcal{T}^{t t}(\mathbf{k}) + \mathcal{T}^{z z}(\mathbf{k}) = \upmu \int_{-\infty}^\infty  \cosh 2 a\eta \exp \left( \frac{i}{a}\mathcal{K} \right)d\eta.
\label{eq:fourier_T_t_t_plus_T_z_z}
\end{equation}
Thus, the contribution of the string to the emission probability $\mathcal{P}_{\mathbf{k}_{\bot}}^{\mathrm{M}},$ given by Eq.~\eqref{eq:transition_probability_Minkowski2}, vanishes. (Recall that this was the case with
the interaction rate computed with 
the mode function $h_{\mu\nu}^{(\omega,\mathbf{k}_\bot)}$ in the Rindler frame.) Furthermore, the component $\mathcal{T}^{t z}(\mathbf{k})$ is given by 
\begin{equation}
\mathcal{T}^{t z}(\mathbf{k}) = \frac{\upmu}{2} \int_{-\infty}^\infty \sinh 2 a \eta \exp \left( \frac{i}{a}\mathcal{K} \right)d\eta,
\label{eq:fourier_T_t_z}
\end{equation}
which has no contribution from the extended string.

To evaluate the integral in Eq.~\eqref{eq:transition_probability_Minkowski2} we ``boost back'' the momentum variables (see Ref.~\cite{higuchi_1992}) by defining
\begin{eqnarray}
\label{eq:kz_linha}
k'_{z} &=& k_z \cosh a \tilde{\eta} - k \sinh a \tilde{\eta}, \\
k' &=& k \cosh a \tilde{\eta} - k_z \sinh a \tilde{\eta}.
\label{eq:k_linha}
\end{eqnarray}
Then, the integrand of Eq.~\eqref{eq:transition_probability_Minkowski2} takes the following form:
\begin{equation}
\frac{1}{2}\abs{\mathcal{T}^{t t}+\mathcal{T}^{z z}}^2 - 2 \abs{\mathcal{T}^{t z}}^2 = \frac{\upmu^2}{2}
\int_{-\infty}^\infty\left[ \int_{-\infty}^\infty \cosh \left[ 2 a (\eta'-\eta'') \right] \exp \left\{ \frac{2i}{a}\left[ k' \sinh \left(\frac{a}{2}(\eta'-\eta'') \right) \right] \right\}d\eta'' \right] d\eta'.
\label{eq:integrand_2}
\end{equation}
By changing variables as $\tilde{\eta} \equiv (\eta'+\eta'')/2$ and $\sigma \equiv \eta' - \eta''$ we can factor out the infinite
time $T_0=\int_{-\infty}^\infty d\tilde{\eta},$ which makes the integral in~\eqref{eq:integrand_2} infinite.
The differential response rate in the inertial frame can be expressed as
\begin{eqnarray}
\mathcal{R}_{\mathbf{k}_{\bot}}^{\mathrm{M}} &=& \frac{\mathcal{P}_{\mathbf{k}_{\bot}}^{M}}{T_0} \nonumber \\
&=& \upmu^2 \int_{-\infty}^\infty \left[ \int_{-\infty}^\infty\cosh 2 a \sigma \exp \left\{ \frac{2i}{a}\left[ k' \sinh \left(\frac{a}{2} \sigma \right) \right] \right\} d\sigma \right]\frac{dk'_z}{4 k' (2\pi)^3}.
\label{eq:response_rate_Mink_0}
\end{eqnarray}
We introduce a convergence factor by letting 
$\sigma \mapsto \sigma + 2i \epsilon,$ where $\epsilon$ is a positive real number, to make the
$\sigma$-integral convergent.  Thus,
\begin{equation}
\mathcal{R}_{\mathbf{k}_{\bot}}^{\mathrm{M}} = \upmu^2 \int_{-\infty}^\infty  \left\{\int_{-\infty}^{\infty}  \cosh 2 a \sigma \exp \left[ \frac{i k'}{a}\left(e^{i a \epsilon} e^{a \sigma/2} - e^{-i a \epsilon} e^{-a \sigma/2} \right) \right]d\sigma  \right\}\frac{dk'_z}{4 k' (2\pi)^3}.
\label{eq:response_rate_Mink_1}
\end{equation}
We take the limit $\epsilon\to 0$ at the end of the calculation.  Now, we introduce the change of variables:
\begin{equation}
s_{\pm} = \frac{k'+ k'_z}{k_{\bot}}e^{\pm a \sigma/2}.
\label{eq:chang_varia_s}
\end{equation} 
Then we find
\begin{equation}
\mathcal{R}_{\mathbf{k}_{\bot}}^{\mathrm{M}} = \frac{\upmu^2}{64\pi^3 a} \int_{0}^{\infty} \left\{ \int_{0}^{\infty}\left(s_+s_- + \frac{1}{s_+^3s_{-}^3} \right) \exp \left[ \frac{i \mu}{2} \left(s_+ - \frac{\beta^2}{s_+} \right)\right]\exp \left[\frac{i \mu}{2} \left(s_- - \frac{\beta^2}{s_-} \right)\right]ds_+\right\}ds_{-},
\label{eq:response_rate_Mink_2}
\end{equation}
where $\mu \equiv k_\perp e^{i a \epsilon}/a$ and $\beta \equiv e^{-i a \epsilon}.$ By using the 
formula~\cite[Eq.~3.471.10]{gradshteyn_2014},
\begin{equation}
\int_{0}^{\infty}ds s^{\nu -1 } \exp\left[\frac{i \mu}{2} \left( s-\frac{\beta^2}{s}\right) \right] = 2 \beta^{\nu} e^{i \nu \pi/2} K_{\nu}(\beta \mu),
\label{eq:BesselK}
\end{equation}
\end{widetext}
for $\Im \mu>0$ and $\Im(\beta^2 \mu)<0,$
we obtain
\begin{equation}
\mathcal{R}_{\mathbf{k}_{\bot}}^{\mathrm{M}} = \frac{\upmu^2}{8\pi^3 a} \abs{K_{2}\left(\frac{k_{\bot}}{a}\right)}^2,
\label{eq:response_rate_Mink}
\end{equation}
which is equal to the differential response rate $\mathcal{R}_{\mathbf{k}_\bot}$ in 
Eq.~\eqref{eq:response_rate}.

\section{Final Remarks}
\label{sec:remarks}
Our work in this paper verifies that the equivalence between the response rates in inertial and co-accelerated frames, observed for a uniform accelerating classical pointlike source interacting with scalar~\cite{ren_1993} or electromagnetic perturbations~\cite{higuchi_1992R,higuchi_1992}, also holds for gravitational perturbations at the lowest order in perturbation theory. For gravitational perturbations, the classical source is a conserved stress-energy tensor describing a point mass 
moving in a hyperbolic trajectory associated with $\xi=0$ in Rindler coordinates with metric~\eqref{eq:line_element} and attached to a string extended from $\xi=0$ to $\infty$. This string can be viewed as the agent driving the acceleration of the point mass, although it acts also as a source of stress energy. 
In the scalar and electromagnetic settings the agent driving the acceleration is there but does not couple with scalar or electromagnetic perturbations provided if it is not charged.

It is interesting to note that the response rate of the classical system coupled with a massless field with spin $s$ at the lowest order (where $s$ takes the values of $0$, $1$, and $2$ for scalar, electromagnetic, and gravitational perturbations, respectively) in the Minkowski vacuum state, can be written in the following suggestive form:
\begin{equation}
\mathcal{R}_{s, \mathbf{k}_{\bot}} = \frac{q^2}{4 \pi^3 s!a} \abs{K_{s}\left(\frac{k_{\bot}}{a}\right)}^2,
\label{eq:response_rate_spin_s}
\end{equation}
where $q$ is the coupling constant, which can be interpreted as the charge or mass of the pointlike object. However, there is a puzzling
feature in the gravitational case ($s=2$) studied in this paper.  
Since $K_2(z) \sim z^{-2}$ for small $z$, we have $\mathcal{R}_{2,\mathbf{k}_{\bot}}\sim k_\bot^{-4}$ for small $k_\bot$.  This implies
that the power of the gravitational radiation, which is bounded below by the integral of 
$k_\perp\mathcal{R}_{2,\mathbf{k}_{\bot}}\sim k_\bot^{-3}$ over $\mathbf{k}_\perp$,
is infrared divergent because $d^2\mathbf{k}_{\bot}\sim k_{\bot} dk_{\bot}$.  The physical origin of 
this infrared divergence needs to be investigated.

\bigskip

\noindent
\textit{Note added.}  Recently, we have become aware of a preprint~\cite{landulfo_2023} that contains some of the
results presented in this paper.

\begin{acknowledgments}
The authors thank Funda\c{c}\~ao Amaz\^onia de Amparo a Estudos e Pesquisas (FAPESPA),  Conselho Nacional de Desenvolvimento Cient\'ifico e Tecnol\'ogico (CNPq), and Coordena\c{c}\~ao de Aperfei\c{c}oamento de Pessoal de N\'{\i}vel Superior (Capes) - Finance Code 001, in Brazil, for partial financial support.
J.B. and L.C. thank the University of York in England and University of Aveiro in Portugal, respectively, for the kind hospitality during the completion of this work.
This work has further been supported by the European Union's Horizon 2020 research and innovation (RISE) program H2020-MSCA-RISE-2017 Grant No. FunFiCO-777740 and by the European Horizon Europe staff exchange (SE) program HORIZON-MSCA-2021-SE-01 Grant No. NewFunFiCO-101086251.
\end{acknowledgments}

\appendix
\section{\uppercase{The response rate with $\eta-$dependent stress-energy tensor}}
\label{sec:appendix_C}

In this appendix we find the response rate of the gravitational field to a conserved stress-energy tensor which is adiabatically turned
on and off in order to dispense with the formal argument using the infinite interaction time.
We first construct a stress-energy tensor $\widetilde{T}^{\mu\nu}$ with nonzero components $\widetilde{T}^{\eta\eta}$, 
$\widetilde{T}^{\eta\xi},$ and $\widetilde{T}^{\xi\xi}$ which depend on $\eta$ as well as $\xi$ in Rindler coordinates.

The conservation equation $\nabla_{\mu}\widetilde{T}^{\eta\mu}=0$ implies
\begin{equation}
e^{-4a\xi}\partial_{\xi} \left( e^{4a\xi} \widetilde{T}^{\eta\xi} \right) = - \partial_{\eta} \widetilde{T}^{\eta \eta}.
\label{eq:T_eta_xi_component}
\end{equation}
We choose $\widetilde{T}^{\eta \eta}$ to be the same as before but 
multiplied by a function $g(\eta),$ i.e.,
\begin{equation}
\widetilde{T}^{\eta \eta}= g(\eta) T^{\eta \eta},
\label{eq:regul_T_eta_eta}
\end{equation}
where $T^{\eta \eta}$ is given by Eq.~\eqref{eq:Rindler_T_eta_eta}. 
With this assumption we find the following solution to Eq.~\eqref{eq:T_eta_xi_component}:
\begin{equation}
\widetilde{T}^{\eta \xi} = - \frac{\upmu}{2}g'(\eta) \left(e^{-4 a \xi} + e^{-2 a \xi} \right) \theta(\xi) \delta^{(2)}(\mathbf{x}_{\bot}).
\label{eq:regul_T_xi_eta}
\end{equation}
Then, the equation $\nabla_{\mu}\widetilde{T}^{\mu \xi}=0$ is satisfied by the following $\xi\xi$-component:
\begin{equation}
\widetilde{T}^{\xi \xi} = g(\eta) T^{\xi \xi} - \frac{\upmu}{2 a}g''(\eta)\left(e^{- 4 a \xi} - e^{- 2 a \xi} \right)\theta(\xi) \delta^{(2)}(\mathbf{x}_{\bot}),
\label{eq:regul_T_xi_xi}
\end{equation}
where $T^{\xi \xi}$ is given by Eq.~\eqref{eq:Rindler_T_xi_xi}. If we set $g(\eta)=1,$ we recover the 
$\eta$-independent stress-energy tensor found in Sec.~\ref{sec:stress-energy}. 
We shall find the interaction probability for this stress-energy tensor with
a general compactly supported and smooth function $g(\eta)$ 
in the inertial and co-accelerated frames.

We start with the inertial frame. 
We define from the stress-energy tensor 
$\widetilde{T}^{\mu\nu}$ given by
Eqs.~\eqref{eq:regul_T_eta_eta}--\eqref{eq:regul_T_xi_xi} its Fourier transform 
$\widetilde{\mathcal{T}}^{\mu\nu}(\mathbf{k})$ exactly as in
Eq.~\eqref{eq:fourier_T} and then Eq.~\eqref{eq:fourier_T_coord_eta_xi} in Rindler coordinates.   We also define the rapidity $\vartheta$ by
\begin{equation}
    \vartheta  \equiv \frac{1}{2a}\log \frac{k+k_z}{k-k_z}. \label{eq:rapidity}
\end{equation}
Then the Fourier transform of $\widetilde{T}^{\mu\nu}(x)$ is given by
\begin{eqnarray}
&& \widetilde{\mathcal{T}}^{\mu \nu}(\mathbf{k})\nonumber \\
&& = \int_{-\infty}^\infty \left[\int_{-\infty}^\infty e^{2 a \xi -i(k_\perp/a)e^{a\xi}\sinh a(\vartheta-\eta)}
\widetilde{T}^{\mu \nu}(x)d\xi\right]d\eta. \nonumber\\
\label{eq:fourier_tildeT_coord_eta_xi}
\end{eqnarray}
The integral over $\eta$ is effectively in a finite interval because the function $g(\eta)$ is compactly supported by assumption.  The integrand grows
in general for large $\xi$ but we make the integral over 
$\xi$ absolutely convergent (unless $k_\bot=0$) by introducing a convergence factor by replacing 
$\sinh a(\vartheta-\eta)$ by $\sinh a(\vartheta-\eta) - i\epsilon$, $\epsilon > 0$, and taking the limit $\epsilon\to 0$ in the end.

Going back to the Minkowski coordinates and integrating by parts, one finds $k_\nu \widetilde{\mathcal{T}}^{\mu\nu} = 0$, from the conservation equation
$\nabla_\nu \widetilde{T}^{\mu\nu} = 0$.  This equation implies $k \widetilde{\mathcal{T}}^{tt} = k_z \widetilde{\mathcal{T}}^{tz}$ and
$k \widetilde{\mathcal{T}}^{tz} = k_z \widetilde{\mathcal{T}}^{zz}$.  These equations can be used to show that
\begin{equation}
    \frac{1}{2}|\widetilde{\mathcal{T}}^{tt} + \widetilde{\mathcal{T}}^{zz}|^2 - 2|\widetilde{\mathcal{T}}^{tz}|^2 = \frac{1}{2}|\widetilde{\mathcal{T}}|^2,
\end{equation}
where $\widetilde{\mathcal{T}} = \widetilde{\mathcal{T}}^{tt} - \widetilde{\mathcal{T}}^{zz}$.  Hence, the differential interaction probability with $\mathbf{k}_\bot$ fixed is
\begin{equation}\label{eq:general-g-response}
    \widetilde{\mathcal{P}}^{\mathrm{M}}_{\mathbf{k}_\bot} = \frac{1}{2}\int \frac{dk_z}{(2\pi)^32k}|\widetilde{\mathcal{T}}^{\mathrm{M}}(\mathbf{k})|^2,
\end{equation}
where
\begin{eqnarray}
\widetilde{\mathcal{T}}^{\mathrm{M}}(\mathbf{k})
& = & \int_{-\infty}^\infty d\eta \int_{-\infty}^\infty d\xi\notag \\
&&  \times e^{2 a \xi -i(k_\perp/a)e^{a\xi}\sinh a(\vartheta-\eta)}
\widetilde{T}_{(\mathrm{R})}(\eta,\xi). \label{eq:general-g-trace}
\end{eqnarray}
The trace $\widetilde{T}_{(\mathrm{R})}(\eta,\xi) = \widetilde{T}_{(\mathrm{R})}^\mu{}_\mu(\eta,\xi)$ can be found from Eqs.~\eqref{eq:regul_T_eta_eta} and \eqref{eq:regul_T_xi_xi},
with $T^{\mu\nu}$ given by Eqs.~\eqref{eq:Rindler_T_eta_eta} and \eqref{eq:Rindler_T_xi_xi}, as
\begin{eqnarray}
    \widetilde{T}_{(\mathrm{R})}(\eta,\xi) &  = & \upmu g(\eta)\left[\delta(\xi) + 2a \theta(\xi)\right] \nonumber \\ 
   &&     + \frac{\upmu}{2a}g''(\eta)(e^{-2a\xi} - 1)\theta(\xi).  \label{eq:trace-of-T}
\end{eqnarray}
By substituting this equation into Eq.~\eqref{eq:general-g-trace} we find
\begin{eqnarray}
    \widetilde{\mathcal{T}}^{\mathrm{M}}(\mathbf{k}) & = & \upmu\int_{-\infty}^\infty F(\eta,\vartheta)d\eta, \label{eq:TMk-by-Fetavartheta}
\end{eqnarray}
where
\begin{widetext}
\begin{eqnarray}
    F(\eta,\vartheta) & = & g(\eta)\left[ e^{-i(k_\bot/a)\sinh a(\vartheta-\eta)} + \frac{2a^2}{k_\bot^2}\int_{k_\bot/a}^\infty 
    e^{-iz[\sinh a(\vartheta-\eta)-i\epsilon]}zdz\right] 
    \nonumber \\
    & & +\frac{1}{2}g''(\eta)\left[ \frac{1}{a^2}\int_{k_\bot/a}^\infty e^{-iz[\sinh a(\vartheta-\eta)-i\epsilon]}\frac{dz}{z} - 
    \frac{1}{k_\bot^2}\int_{k_\bot/a}^\infty e^{-iz[\sinh a(\vartheta-\eta)-i\epsilon]}zdz\right].
\end{eqnarray}

Next we derive the differential interaction probability with $\mathbf{k}_\bot$ fixed in the right Rindler wedge and show that it equals $\widetilde{\mathcal{P}}^{\mathrm{M}}_{\mathbf{k}_\bot}$ given by 
Eq.~\eqref{eq:general-g-response}.  The emission and absorption
amplitudes corresponding to Eqs.~\eqref{eq:amplitude_R_emission2} and \eqref{eq:amplitude_R_absorption2}
can be found using the stress-energy tensor given by Eqs.~\eqref{eq:regul_T_eta_eta}--\eqref{eq:regul_T_xi_xi} and the mode functions given by Eqs.~\eqref{new-field} and \eqref{new-field-transverse} as
\begin{align}
    \widetilde{\mathcal{A}}^{(\omega,\mathbf{k}_\bot)}_{\mathrm{em}}
    & = -\sqrt{2}\int \widetilde{T}_{(\mathrm{R})}(\eta,\xi)\phi^{(\omega,k_\bot)}(\xi)e^{i\omega\eta - i\mathbf{k}_\bot\cdot\mathbf{x}_\bot}\sqrt{-g}\,d^4 x,
\end{align}
where the trace of the stress-energy tensor, $\widetilde{T}_{(\mathrm{R})}(\eta,\xi)$, and the
functions $\phi^{(\omega,k_\bot)}(\xi)$ are given by Eqs.~\eqref{eq:trace-of-T} and 
\eqref{eq:solution_phi},  respectively.  Thus, we find the one-graviton emission amplitude in the 
right Rindler wedge as
\begin{align}
    \widetilde{\mathcal{A}}^{(\omega,\mathbf{k}_\bot)}_{\mathrm{em}} & =
    - \sqrt{2}\upmu \int_{-\infty}^\infty \left[ g(\eta)\phi^{(\omega,k_\bot)}(0)
    + 2a\int_{-\infty}^\infty \phi^{(\omega,k_\bot)}(\xi)e^{2a\xi}d\xi \right]e^{i\omega\eta}d\eta \notag \\
    & \quad - \frac{\upmu}{\sqrt{2}a}\int_{-\infty}^\infty\left[ \int_{-\infty}^\infty 
    g''(\eta)\phi^{(\omega,k_\bot)}(\xi)(1-e^{2a\xi}) d\xi \right]e^{i\omega\eta}d\eta,
    \label{eq:Aem-in-Rindler}
\end{align}
\end{widetext}
and the one-graviton absorption amplitude as
\begin{align}
    \widetilde{\mathcal{A}}^{(\omega,\mathbf{k}_\bot)}_{\mathrm{abs}}
    = \overline{\widetilde{\mathcal{A}}^{(\omega,-\mathbf{k}_\bot)}_{\mathrm{em}}}.
\end{align}
Then, the interaction probability is the sum of the (spontaneous and induced) emission probability
and the absorption probability in the FDU thermal bath of gravitons:
\begin{align}
    \widetilde{\mathcal{P}}^{\mathrm{R}} = 
    \int \widetilde{\mathcal{P}}_{\mathbf{k}_\bot}^{\mathrm{R}}\,d^2\mathbf{k}_\bot,
\end{align}
where
\begin{align}
    \widetilde{\mathcal{P}}_{\mathbf{k}_\bot}^{\mathrm{R}} 
    & = \int_{0}^\infty \left[ |\widetilde{\mathcal{A}}^{(\omega,\mathbf{k}_\bot)}_{\mathrm{em}}|^2[1 + n(\omega)] + |\widetilde{\mathcal{A}}^{(\omega,-\mathbf{k}_\bot)}_{\mathrm{abs}}|^2 n(\omega) \right]d\omega  \notag \\
    & = \int_0^\infty \left[ \frac{|\widetilde{\mathcal{A}}^{(\omega,\mathbf{k}_\bot)}_{\mathrm{em}}|^2}{1 - e^{-2\pi\omega/a}} + \frac{|\widetilde{\mathcal{A}}^{(\omega,-\mathbf{k}_\bot)}_{\mathrm{abs}}|^2}{e^{2\pi\omega/a} - 1}\right]d\omega.  \label{eq:emission-plus-absorption}
\end{align}

Now, we show that Eq.~\eqref{eq:emission-plus-absorption} agrees with the emission probability 
$\widetilde{\mathcal{P}}^{\mathrm{M}}_{\mathbf{k}_\bot}$ given by Eq.~\eqref{eq:general-g-response}, which was found in the standard Minkowski-spacetime calculation.
An important fact for this purpose is that, for scalar field theory, 
there is a complete set of mode functions which are positive-frequency in the right Rindler wedge and negative-frequency in the left Rindler wedge, or vice versa, 
with respect to the Killing vector $\partial_\eta$ and are positive-frequency with respect to the Minkowski Killing vector $\partial_t$.  It is easy to see that
the same is true for the even-parity sector of the gravitational field, because the even-parity mode functions are scalar mode functions times a constant tensor.
Let us explain this fact in more detail.
We choose the following mode functions defined over the whole Minkowski spacetime:
\begin{equation}
\widetilde{h}^{(\mathrm{R};\omega,\mathbf{k}_\bot)}_{\mu\nu} \equiv \begin{cases}  \widetilde{h}^{(\omega,\mathbf{k}_\bot)}_{\mu\nu} & \textrm{in the right Rindler wedge},\\
0 & \textrm{in the left Rindler wedge}. \end{cases}
\end{equation}
If we define the Rindler coordinate system for the left Rindler wedge, $z < -|t|$, by letting $t=a^{-1}e^{a\xi}\sinh a\eta$ and $z = - a^{-1}e^{a\xi}\cosh a\eta$, 
the functions $\widetilde{h}^{(\omega,\mathbf{k}_\bot)}_{\mu\nu}(\eta,\xi,\mathbf{x}_\bot)$ define mode functions in the left Rindler wedge, which are of
positive frequency with respect to the Killing vector $\partial_\eta$ there.  Let us now define
\begin{equation}
\widetilde{h}^{(\mathrm{L};\omega,\mathbf{k}_\bot)}_{\mu\nu} \equiv \begin{cases}  
0 & \textrm{in the right Rindler wedge}, \\
\widetilde{h}^{(\omega,\mathbf{k}_\bot)}_{\mu\nu} & \textrm{in the left Rindler wedge}.
\end{cases}
\end{equation}
Then, exactly as in the scalar case~\cite{unruh_1976}, the following mode functions, which we call the Unruh modes,
are of positive frequency with respect to the Minkowski Killing vector $\partial_t$:
\begin{eqnarray}
\widetilde{h}^{(-;\omega,\mathbf{k}_\bot)}_{\mu\nu}
& = & \frac{\widetilde{h}^{(\mathrm{R};\omega,\mathbf{k}_\bot)}_{\mu\nu}}{\sqrt{ 1- e^{-2\pi\omega/a}}}
+ \frac{\overline{\widetilde{h}^{(\mathrm{L};\omega,-\mathbf{k}_\bot)}_{\mu\nu}}}{\sqrt{e^{2\pi\omega/a} -1 }},\\
\widetilde{h}^{(+;\omega,\mathbf{k}_\bot)}_{\mu\nu}
& = & \frac{\overline{\widetilde{h}^{(\mathrm{R};\omega,-\mathbf{k}_\bot)}_{\mu\nu}}}{\sqrt{ e^{2\pi\omega/a}-1}}
+ \frac{\widetilde{h}^{(\mathrm{L};\omega,\mathbf{k}_\bot)}_{\mu\nu}}{\sqrt{1 - e^{-2\pi\omega/a}}}.
\end{eqnarray}

Now, the quantum graviton field for the even-parity sector can then be expanded in terms of the Unruh modes:
\begin{eqnarray}
    \widehat{h}_{\mu\nu}^{(\textrm{even})}
    & = & \int_{0}^\infty d\omega \int d^2\mathbf{k}_\bot \left[ \widetilde{h}_{\mu\nu}^{(-;\omega,\mathbf{k}_\bot)}\hat{a}^{(-)}_{(\omega,\mathbf{k}_\bot)} \right. \nonumber \\
    & &  \qquad + \left.
    \widetilde{h}_{\mu\nu}^{(+;\omega,\mathbf{k}_\bot)}\hat{a}^{(+)}_{(\omega,\mathbf{k}_\bot)} +
    \textrm{H.c.} \right]. \label{eq:tensor_mode_operator_unruh}
\end{eqnarray}
The inner product for the Unruh modes,
$\widetilde{h}_{\mu\nu}^{(-;\omega,\mathbf{k}_\bot)}$ and
$\widetilde{h}_{\mu\nu}^{(+;\omega,\mathbf{k}_\bot)}$,
can be found using the inner product for the Rindler modes, 
$\widetilde{h}_{\mu\nu}^{(\mathrm{R};\omega,\mathbf{k}_\bot)}$ and
$\widetilde{h}_{\mu\nu}^{(\mathrm{L};\omega,\mathbf{k}_\bot)}$,
given by Eq.~\eqref{eq:inner-product-even-modes}.  One finds that the Unruh modes defined
here are orthogonal to one another and
\begin{eqnarray}
    \langle \widetilde{h}^{(-;\omega,\mathbf{k}_\bot)},\widetilde{h}^{(-;\omega',\mathbf{k}_\bot')}\rangle_{\textrm{dD}}
    & = & \langle \widetilde{h}^{(+;\omega,\mathbf{k}_\bot)},\widetilde{h}^{(+;\omega',\mathbf{k}_\bot')}\rangle_{\textrm{dD}} \nonumber \\
    & = & \delta(\omega-\omega')\delta^{(2)}(\mathbf{k}_\bot - \mathbf{k}'_\bot).\nonumber \\ \label{eq:inner-product-Unruh-modes}
\end{eqnarray}
This implies that the annihilation operators, $\hat{a}^{(-)}_{(\omega,\mathbf{k}_\bot)}$ and $\hat{a}^{(+)}_{(\omega,\mathbf{k}_\bot)}$, and the creation operators,
$\hat{a}^{(-)\dagger}_{(\omega,\mathbf{k}_\bot)}$ and
$\hat{a}^{(+)\dagger}_{(\omega,\mathbf{k}_\bot)}$,
satisfy the standard commutation relations,
\begin{eqnarray}
\left[ \hat{a}^{(-)}_{(\omega,\mathbf{k}_\bot)}, \hat{a}^{(-)\dagger}_{(\omega',\mathbf{k}'_\bot)}\right]
& = &\left[ \hat{a}^{(+)}_{(\omega,\mathbf{k}_\bot)}, \hat{a}^{(+)\dagger}_{(\omega',\mathbf{k}_\bot')}\right]\nonumber \\
& = & \delta(\omega-\omega')\delta^{(2)}(\mathbf{k}_\bot - \mathbf{k}_\bot'),\nonumber \\
\end{eqnarray}
with all other commutators vanishing.  Since the mode functions for the operators
$\hat{a}^{(-)}_{(\omega,\mathbf{k}_\bot)}$ and
$\hat{a}^{(+)}_{(\omega,\mathbf{k}_\bot)}$ are positive-frequency with respect to the
Minkowski Killing vector $\partial_t$, the Minkowski vacuum state $\ket{0_{\mathrm{M}}}$ is annihilated
by these operators: $\hat{a}^{(-)}_{(\omega,\mathbf{k}_\bot)}\ket{0_{\mathrm{M}}} =
\hat{a}^{(+)}_{(\omega,\mathbf{k}_\bot)}\ket{0_{\mathrm{M}}} = 0$.

The one-graviton final state due to the stress-energy tensor $\widetilde{T}^{\mu\nu}(x)$ 
can be found as
\begin{eqnarray}
    \ket{\textrm{1g}} & = & i \int \widetilde{T}^{\mu\nu}(x)\widehat{h}_{\mu\nu}^{(\textrm{even})}(x)\sqrt{-g}\,d^4x\,\ket{0_\mathrm{M}} \nonumber   \\
    & = & i\int_0^\infty d\omega \int d^2\mathbf{k}_\bot   \left[ \frac{\widetilde{\mathcal{A}}^{(\omega,\mathbf{k}_\bot)}_{\mathrm{em}}}{\sqrt{1-e^{-2\pi\omega/a}}}\hat{a}^{(-)\dagger}_{(\omega,\mathbf{k}_\bot)}\right. \nonumber \\
    & & \qquad \quad \left. + \frac{\widetilde{\mathcal{A}}^{(\omega,-\mathbf{k}_\bot)}_{\mathrm{abs}}}{\sqrt{e^{2\pi\omega/a}-1}} 
    \hat{a}^{(+)\dagger}_{(\omega,\mathbf{k}_\bot)}\right]\ket{0_{\mathrm{M}}}.
    \label{eq:1g-expanded}
\end{eqnarray}
Then we find
\begin{align}
    \langle \textrm{1g}|\textrm{1g}\rangle = \widetilde{\mathcal{P}}_{\mathbf{k}_\bot}^{\mathrm{R}},
    \label{eq:1g-1g}
\end{align}
where $\widetilde{\mathcal{P}}_{\mathbf{k}_\bot}^{\mathrm{R}}$ is given by 
Eq.~\eqref{eq:emission-plus-absorption}.  Thus, one can show that 
$\widetilde{\mathcal{P}}_{\mathbf{k}_\bot}^{\mathrm{R}} = \widetilde{\mathcal{P}}_{\mathbf{k}_\bot}^{\mathrm{M}}$ by showing that $\langle \textrm{1g}|\textrm{1g}\rangle = \widetilde{\mathcal{P}}_{\mathbf{k}_\bot}^{\mathrm{M}}$.  

To show that $\langle \textrm{1g}|\textrm{1g}\rangle = \widetilde{\mathcal{P}}_{\mathbf{k}_\bot}^{\mathrm{M}}$ we need to express the creation operators
$\hat{a}^{(-)\dagger}_{(\omega,\mathbf{k}_\bot)}$ and $\hat{a}^{(+)\dagger}_{(\omega,\mathbf{k}_\bot)}$ in terms of the creation operators for the momentum eigenstates
in Minkowski spacetime.
We define the following mode function with momentum $\mathbf{k}$ in Minkowski spacetime:
\begin{align}
    \widetilde{h}_{\mu\nu}^{(\mathbf{k})}(x) 
    = - \frac{1}{\sqrt{2}}\left[ g_{\mu\nu} + 2q_{\mu\nu}(\mathbf{k}_\bot)\right]
    \varphi^{(\mathbf{k})}(x),
\end{align}
where
\begin{align}
    \varphi^{(\mathbf{k})}(x) = \frac{1}{\sqrt{(2\pi)^32k}}e^{-ikt + i\mathbf{k}\cdot\mathbf{x}}.
\end{align}
Comparing this definition with Eq.~\eqref{new-field} one can readily see that the relationship between
$\widetilde{h}^{(\omega,\mathbf{k}_\bot)}_{\mu\nu}(\eta,\xi,\mathbf{x}_\bot)$ and 
$\widetilde{h}_{\mu\nu}^{(\mathbf{k})}(t,\mathbf{x})$ is the same as
that between $\varphi^{(\omega,\mathbf{k}_\bot)}(\eta,\xi,\mathbf{x}_\bot)$ and $\varphi^{(\mathbf{k})}(t,\mathbf{x})$.  
This implies that
the relationship between the Unruh modes $\widetilde{h}^{(\pm;\omega,\mathbf{k}_\bot)}_{\mu\nu}$ and
the Minkowski modes $\widetilde{h}^{(\mathbf{k})}_{\mu\nu}$ is identical to that between the Unruh modes
and the Minkowski modes for the scalar field~\cite{crispino_2008}.  Thus,
\begin{align}
    \widetilde{h}_{\mu\nu}^{(\pm;\omega,\mathbf{k}_\bot)}
    = \int_{-\infty}^\infty e^{\pm i\omega\vartheta(k_z)}
    \widetilde{h}_{\mu\nu}^{(\mathbf{k})}\frac{dk_z}{\sqrt{2\pi ak}},
\end{align}
where the rapidity $\vartheta(k_z)$ is defined by Eq.~\eqref{eq:rapidity}.  This relation can be
inverted as
\begin{align}
    \widetilde{h}_{\mu\nu}^{(\mathbf{k})} & = 
    \frac{1}{\sqrt{2\pi ak}}\int_0^\infty \left[ e^{i\omega\vartheta(k_z)}
    \widetilde{h}_{\mu\nu}^{(-;\omega,\mathbf{k}_\bot)} \right. \notag \\ 
    & \qquad \qquad \qquad \qquad \left. + e^{-i\omega\vartheta(k_z)}
    \widetilde{h}_{\mu\nu}^{(+;\omega,\mathbf{k}_\bot)}\right]d\omega. \label{eq:hk-expansion}
\end{align}
The field describing the even-parity sector can be expanded as
\begin{align}
    \widehat{h}_{\mu\nu}^{(\mathrm{even})}(x)
    = \int \left[ \widetilde{h}_{\mu\nu}^{(\mathbf{k})}(x)\hat{a}_{\mathbf{k}}
    + \overline{\widetilde{h}_{\mu\nu}^{(\mathbf{k})}(x)}\hat{a}^\dagger_{\mathbf{k}}\right]
    d^3\mathbf{k}, \label{eq:alternative-even-expansion}
\end{align}
where the operators $\hat{a}_{\mathbf{k}}$ and $\hat{a}^\dagger_{\mathbf{k}}$ satisfy
\begin{align}
    \left[\hat{a}_{\mathbf{k}},\hat{a}_{\mathbf{k}'}^{\dagger}\right] = \delta^{(3)}(\mathbf{k}-\mathbf{k}').
\end{align}
By substituting Eq.~\eqref{eq:hk-expansion} into Eq.~\eqref{eq:alternative-even-expansion} and 
comparing the coefficients of $\widetilde{h}_{\mu\nu}^{(\pm;\omega,\mathbf{k}_\bot)}$ with the expansion~\eqref{eq:tensor_mode_operator_unruh}, we find
\begin{align}\label{eq:Rind-in-terms-of-Mink}
     a^{(\pm)\dagger}_{(\omega,\mathbf{k}_\bot)} & = \int_{-\infty}^\infty \frac{dk_z}{\sqrt{2\pi ak}}\,e^{\pm i\omega\vartheta}\, a^\dagger_{\mathbf{k}}.
\end{align}

We substitute Eq.~\eqref{eq:Rind-in-terms-of-Mink} into Eq.~\eqref{eq:1g-expanded} and find
\begin{align}
    \ket{1\mathrm{g}} & = i\int \frac{d^3\mathbf{k}}{\sqrt{(2\pi)^3 2k}}\, \mathcal{A}^{(\mathrm{R})}(\mathbf{k}) \hat{a}_\mathbf{k}^\dagger\ket{0_\mathrm{M}}, \label{eq:1g-in-rindler}
\end{align}
where 
\begin{align}
    \mathcal{A}^{(\mathrm{R})}(\mathbf{k})
    = \sqrt{\frac{8\pi^2}{a}}\int_{-\infty}^\infty \frac{e^{-i\omega\vartheta(k_z)}
    \widetilde{\mathcal{A}}_{\mathrm{em}}^{(\omega,\mathbf{k}_\bot)}}{\sqrt{1-e^{-2\pi\omega/a}}}d\omega. \label{eq:AR-Aem}
\end{align}
Here, we have used the fact that 
$\widetilde{\mathcal{A}}_{\mathrm{abs}}^{(\omega,-\mathbf{k}_\bot)}/\sqrt{e^{2\pi\omega/a}-1}$ can
be obtained by letting $\omega\to -\omega$ in the expression
$\widetilde{\mathcal{A}}_{\mathrm{em}}^{(\omega,\mathbf{k}_\bot)}/\sqrt{1-e^{-2\pi\omega/a}}$.
Then, Eqs.~\eqref{eq:1g-in-rindler} and \eqref{eq:1g-1g} imply
\begin{align}
    \widetilde{\mathcal{P}}^{\mathrm{R}}_{\mathbf{k}_\bot} = \int \frac{dk_z}{(2\pi)^32k}|\mathcal{A}^{(\mathrm{R})}(\mathbf{k})|^2.
\end{align}

We now show that
\begin{align}
   \sqrt{2}\,\mathcal{A}^{(\mathrm{R})}(\mathbf{k}) = - \widetilde{\mathcal{T}}^{\mathrm{M}}(\mathbf{k}).  \label{eq:equality-R-and-M}
\end{align}
This will establish that $\widetilde{\mathcal{P}}^{\mathrm{R}}_{\mathbf{k}_\bot} = \widetilde{\mathcal{P}}^{\mathrm{M}}_{\mathbf{k}_\bot}$ 
[see Eq.~\eqref{eq:general-g-response}].  First, we find from Eq.~\eqref{eq:solution_phi},
\begin{align}
    \frac{\phi^{(\omega,k_\bot)}(\xi)}{\sqrt{1-e^{-2\pi\omega/a}}}
    & = \frac{e^{\pi\omega/2a}}{\sqrt{8\pi^4 a}}K_{i\omega/a}\left(\frac{k_\bot}{a}e^{a\xi}\right).
\end{align}
By using the identity~\cite[Eq.~6.796]{gradshteyn_2014},
\begin{equation}
 \int_{-\infty}^{+\infty} e^{-i\omega y} e^{\pi\omega/2a}K_{i\omega/a}(z)d\omega 
 = \pi a e^{-iz\sinh ay},
\end{equation}
we find
\begin{align}
    & \int_{-\infty}^\infty \frac{e^{-i\omega(\vartheta-\eta)}\phi^{(\omega,k_\bot)}(\xi)}{\sqrt{1-e^{-2\pi\omega/a}}}d\omega \notag \\
    &    = \sqrt{\frac{a}{8\pi^2}}e^{-i(k_\bot/a)e^{a\xi}\sinh a(\vartheta-\eta)}.
\end{align}
By combining this result with Eqs.~\eqref{eq:AR-Aem} and \eqref{eq:Aem-in-Rindler} we indeed find
Eq.~\eqref{eq:equality-R-and-M}.  This shows that the response rate to the $\eta$-dependent classical stress-energy tensor given by Eqs.~\eqref{eq:regul_T_eta_eta}--\eqref{eq:regul_T_xi_xi} is reproduced
in the Rindler wedge using the Unruh effect.

In the rest of this appendix we show that Eq.~\eqref{eq:general-g-response} gives the emission rate
$\widetilde{\mathcal{R}}_{\mathbf{k}_\bot}$ in Eqs.~\eqref{eq:response_rate} and \eqref{eq:response_rate_Mink} 
if the function $g(\eta)$ is smooth and compactly supported and equals $1$ for most of the time when 
it is nonzero.
First, we find by integration by parts in $\eta$,
\begin{widetext}
\begin{align}
     e^{-i(k_\bot/a)\sinh a(\vartheta-\eta)}g(\eta) 
    & \simeq \frac{a^2}{k_\bot^2}\left[ \frac{3}{\cosh^4 a(\vartheta-\eta)} - \frac{2}{\cosh^2 a(\vartheta-\eta)}\right]g(\eta) 
      - \frac{2a}{k_\bot^2}\frac{\sinh a(\vartheta-\eta)}{\cosh^3 a(\vartheta-\eta)}g'(\eta) - \frac{g''(\eta)}{k_\bot^2 \cosh^2 a(\vartheta-\eta)}
\end{align}
and
\begin{align}
    \frac{2a^2}{k_\bot}g(\eta)\int_{k_\bot/a}^\infty e^{-iz[\sinh a(\vartheta-\eta) - i\epsilon]}zdz
    & \simeq \frac{2a^2}{k_\bot^2}\frac{g(\eta)}{\cosh^2 a(\vartheta-\eta)} + \frac{2ia}{k_\bot^2}\frac{g'(\eta)}{\cosh a(\vartheta-\eta)}
    \int_{k_\bot/a}^\infty e^{-iz[\sinh a(\vartheta-\eta) - i\epsilon]}dz,
\end{align}
where $\simeq$ denotes the equality under the $\eta$-integral.
We use the following identity to make the $z$-integrals manifestly convergent:
\begin{align}
& \int_{-\infty}^{+\infty} G(\eta,\vartheta) \left[\int_{k_\perp/a}^{+\infty}  e^{-iz\sinh a(\vartheta-\eta)}\frac{d z}{z^n}\right] d\eta
=  \frac{i}{a}\int_{-\infty}^{+\infty} \left\{ \frac{d}{d \eta}
\left[ \frac{G(\eta,\vartheta)}{\cosh a(\vartheta-\eta)} \right] 
\int_{k_\perp/a}^{+\infty}  e^{-iz\sinh a(\vartheta-\eta)}\frac{d z}{z^{n+1}}
\right\}d\eta. 
\label{identity1_app}
\end{align}
We define
the differential operator $\mathcal{D}$ on any smooth function $f(\eta,\vartheta)$ by
\begin{align}
    (\mathcal{D}f)(\eta,\vartheta) = \frac{\partial\ }{\partial\eta}\left[ \frac{f(\eta,\vartheta)}{\cosh a(\vartheta-\eta)}\right].
\end{align}
Then we find
\begin{align}
    F(\eta,\vartheta) & \simeq \frac{3a^2}{k_\bot^2}\frac{g(\eta)}{\cosh^4 a(\vartheta-\eta)}
     -\frac{2a}{k_\bot^2}\frac{\sinh a(\vartheta-\eta)}{\cosh^2 a(\vartheta-\eta)}g'(\eta)
      -\frac{g''(\eta)}{k_\bot^2\cosh^2 a(\vartheta-\eta)} \notag \\
      & \quad + i\left[ - \frac{2}{ak_\bot^2}(\mathcal{D}^2 g')(\vartheta,\eta)
      + \frac{1}{2a^3}(\mathcal{D}g'')(\vartheta,\eta) + \frac{1}{2a^3k_\bot^2}(\mathcal{D}^3g'')(\vartheta,\eta)\right]\int_{k_\bot/a}^\infty e^{-iz\sinh a(\vartheta-\eta)}\frac{dz}{z^2}.
      \label{eq:Fetavarthetafinal}
\end{align}
\end{widetext}

We choose the function $g(\eta)$ to be $1$ for $-T_0 < \eta < T_0$, $0$ for $ |\eta| > T_0 + b,$ and decrease (increase) smoothly from
$\eta=T_0$ to $T_0+b$ (from $\eta=-T_0-b$ to $-T_0$).  
In the end we let $T_0\to \infty$ while keeping $b$ unchanged.  
Now, the terms multiplied by $g'(\eta)$ or $g''(\eta)$
in Eq.~\eqref{eq:Fetavarthetafinal} are nonzero only if $\eta\in [-T_0-b,-T_0]\cup [T_0,T_0+b]$. 
Hence, these terms contribute to the $\eta$-integral only if $\vartheta$ is in or near the
set $[-T_0-b,-T_0]\cup [T_0,T_0+b]$ because of the exponential falloff of these terms for large
$|\vartheta-\eta|$.  Then, the contribution from the terms other than the first term in 
Eq.~\eqref{eq:Fetavarthetafinal} to the integral~\eqref{eq:general-g-response} for
the emission probability $\mathcal{P}^{\mathrm{M}}_{\mathbf{k}_\bot}$,
which is an integral over $\vartheta(k_z)$ because $d\vartheta(k_z)= dk_z/ak$,
remains finite if we keep the
shape of the function $g(\eta)$ for $T_0 \leq |\eta| \leq T_0+b$ unchanged as we take the limit
$T_0\to\infty$.  Therefore, we only need to consider the first term in Eq.~\eqref{eq:Fetavarthetafinal}
if we take the $T_0\to \infty$ limit in this manner.

Thus, the differential emission rate with $\mathbf{k}_\bot$ fixed in this limit can be found as
\begin{align}
   \mathcal{R}_{\mathbf{k}_\perp} & = \lim_{T_0\to\infty}\frac{1}{2T_0}\int_{-\infty}^\infty dk_z |\mathcal{A}(\mathbf{k})|^2 \\
   &  = \frac{9\upmu^2 a^5}{32\pi^3 k_\perp^4}\lim_{T_0\to\infty}\frac{1}{2T_0}\int_{-\infty}^\infty d\vartheta  \notag \\
   & \quad \times \left|\int_{-\infty}^\infty
   g(\vartheta-s)\frac{e^{-i(k_\perp/a)\sinh as}}{\cosh^4 as}ds\right|^2,
\end{align}
where we have defined $s\equiv \vartheta - \eta$.
Because of the factor $\cosh^4 as$ in the denominator the integrand is non-negligible only for $|s| \sim 1/a$.  Thus, for large $T_0$
we may replace $g(\vartheta-s)$ by the characteristic function of the interval $[-T_0,T_0]$, 
which is $1$ if $\vartheta$ is in this interval
and $0$ otherwise.  Hence,
\begin{align}
    \mathcal{R}_{\mathbf{k}_\perp} & = \frac{\upmu^2}{8\pi^3 a}
    \left|\int_{-\infty}^\infty\frac{3a^3e^{-i(k_\perp/a)\sinh as}}{2k_\perp^2\cosh^4 as}ds\right|^2.
\end{align}
We recover the differential emission rate given by Eq.~\eqref{eq:response_rate} using Ref.~\cite[Eq.~8.432.5]{gradshteyn_2014}.

\section{\uppercase{Response rate of another conserved stress-energy tensor}}
\label{sec:appendix_E}
To have some more insight into the gravitational radiation from the accelerated point mass, it may be useful to consider 
the same point mass with an extended string with a different stress-energy tensor.
In this appendix, we give the response rate of the gravitational field to the source with the following conserved stress-energy tensor:
\begin{eqnarray}
T^{\eta \eta} & = &\upmu \delta(\xi)
 \delta^{(2)}(\mathbf{x}_{\bot})\label{eq:app_B_T_eta_eta},\\
 T^{\xi \xi} & = & - \upmu a e^{- 3 a \xi} \theta(\xi)
 \delta^{(2)}(\mathbf{x}_{\bot}),\label{eq:app_B_T_xi_xi}
\end{eqnarray}
with all other components vanishing.  This stress-energy tensor is obtained by letting
$F(\xi) = e^{-a\xi}\theta(\xi)$ in Eq.~\eqref{eq:solution_Txixi}.
\begin{widetext}
We can readily express the stress-energy tensor in Minkowski coordinates using Eqs.~\eqref{eq:coord_transform_t} and~\eqref{eq:coord_transform_z}, namely
\begin{eqnarray}
\label{eq:app_B_Mink_T_t_t}
T^{t t } &=& \upmu \left[ \cosh^2 a\eta \delta (\xi) - a e^{-a\xi} \sinh^2 a\eta \theta(\xi) \right] \delta^{(2)}(\mathbf{x}_{\bot}), \\
\label{eq:app_B_Mink_T_t_z}
T^{t z } &=& \upmu \sinh a\eta \cosh a\eta \left[\delta (\xi) - a e^{-a\xi}  \theta(\xi) \right] \delta^{(2)}(\mathbf{x}_{\bot}), \\
\label{eq:app_B_Mink_T_z_z}
T^{z z } &=& \upmu  \left[ \sinh^2 a\eta \delta (\xi) - a e^{-a \xi} \cosh^2 a\eta \theta(\xi) \right] \delta^{(2)}(\mathbf{x}_{\bot}),
\end{eqnarray}
\end{widetext}
with all other components vanishing. 
The stress-energy tensor associated with Eqs.~\eqref{eq:app_B_T_eta_eta} and~\eqref{eq:app_B_T_xi_xi} describes a point mass in a hyperbolic trajectory, with $\xi=0,$ attached to a massless string whose tension is associated with $T^{\xi \xi}.$  We note that this stress-energy tensor does not satisfy the WEC.

The response rate of the gravitational field to the stress-energy tensor given by Eqs.~\eqref{eq:app_B_T_eta_eta}--\eqref{eq:app_B_Mink_T_z_z} in the inertial and co-accelerated frames can be obtained by performing calculations similar to those in Secs.~\ref{sec:response-rate-Rindler} and~\ref{sec:response-rate-Minkowski}. This response rate is given by
\begin{eqnarray}
\mathcal{R}_{\mathbf{k}_{\bot}} & = & \frac{\upmu^2}{8 \pi^3 a} \left[ K_{2}\left( \frac{k_{\bot}}{a}\right) - \frac{a}{k_{\bot}} \int_{k_{\bot}/a}^{\infty} K_{2}\left(z \right) dz\right]^2
\label{eq:app_B_response_rate} \\
& = & \frac{\upmu^2}{8 \pi^3 a} \left[ K_{0}\left( \frac{k_{\bot}}{a}\right) + \frac{a}{k_{\bot}} \int_{k_{\bot}/a}^{\infty} K_{0}\left(z \right) dz\right]^2. \nonumber \\
\label{eq:app_B_response_rate2}
\end{eqnarray}
Equation~\eqref{eq:app_B_response_rate} is found by using the original mode functions given by
Eqs.~\eqref{eq:SYK_normalized_modes_eta_eta_xi_xi}--\eqref{eq:SYK_normalized_modes_i_j}, whereas
Eq.~\eqref{eq:app_B_response_rate2} is found by using the gauge-transformed modes given by
Eqs.~\eqref{new-field}.  These two expressions can directly 
be shown to be equal by using Eqs.~\eqref{eq:K0-K2-equation} and \eqref{eq:diff-eq-for-K0}.  It is
interesting that this response rate diverges such as $k_\bot^{-2}$ 
in the $k_\bot\to 0$ limit, whereas that for
the stress-energy tensor given by Eqs.~\eqref{eq:Rindler_T_eta_eta} and \eqref{eq:Rindler_T_xi_xi} 
found in Secs.~\ref{sec:response-rate-Rindler} and \ref{sec:response-rate-Minkowski} diverges
as $k_\bot^{-4}$ in this limit.

More generally, if we let $F(\xi) = e^{-\beta a \xi} \theta(\xi)$ in Eq.~\eqref{eq:solution_Txixi}, we find
\begin{align}
    \check{T}^{\eta\eta} & =  \upmu\left[ \delta(\xi) + (1-\beta)ae^{-(2+\beta)a\xi}\theta(\xi)\right]
    \delta^{(2)}(\mathbf{x}_\bot),\\
    \check{T}^{\xi\xi} & = - \upmu a e^{-(2+\beta)a\xi}\theta(\xi)\delta^{(2)}(\mathbf{x}_\bot).
\end{align}
Then,
\begin{align}
 \check{T}^\alpha{}_\alpha & = \upmu\left[ \delta(\xi) + (2-\beta)ae^{-\beta a\xi}\theta(\xi)\right]\delta^{(2)}(\mathbf{x}_\bot).
\end{align}
The stress-energy tensor in Sec.~\ref{sec:stress-energy} corresponds to $\beta=0$ and satisfies the WEC. 
If $\beta > 0$, this stress-energy tensor does not satisfy the WEC. On the other hand, if $\beta < 0$, then its trace grows exponentially as a function of $\xi$.
We find the interaction rate for general $\beta$ with $\mathbf{k}_\bot$ fixed as
\begin{align}
    \check{\mathcal{R}}_{\mathbf{k}_{\bot}}
    & = \frac{\upmu^2}{8 \pi^3 a} \left[ K_{0}\left( \frac{k_{\bot}}{a}\right)\right. \notag \\
    & \qquad \left. + (2-\beta)
    \left(\frac{a}{k_{\bot}}\right)^{2-\beta} \int_{k_{\bot}/a}^{\infty} K_{0}\left(z \right)z^{1-\beta}dz\right]^2.
\end{align}

\


\end{document}